\documentclass{aa}
\usepackage{graphicx}
\usepackage{natbib}
\usepackage{amssymb}
\usepackage{amstext}
\usepackage{latexsym} 
\usepackage[sumlimits]{amsmath}

\usepackage{txfonts}
\begin{document}
\title
{Star-forming galaxies in low-redshift clusters:  
Data and integrated galaxy properties }

\author{C.F.~Thomas\inst{1,2}
	 \and C.~Moss\inst{1}
	 \and P.A.~James\inst{1}
	 \and S.M.~Bennett\inst{3}
	 \and A.~Arag\'{o}n-Salamanca\inst{4} 
	 \and M.~Whittle\inst{5}}

	\offprints{C. F. Thomas}

        \institute{Astrophysics Research Institute, 
	Liverpool John Moores University, 
	Birkenhead CH41 1LD\\
	\email{cft@astro.livjm.ac.uk}
	\and Royal Observatory, Greenwich, London SE10 9NF
	\and Institute of Astronomy, Madingley Road, 
	Cambridge CB3 0HA
	\and School of Physics and Astronomy, 
	The University of Nottingham, 
	University Park, Nottingham NG7 2RD
	\and Department of Astronomy, University of 
	  Virginia, Charlottesville, VA 22903, USA}

\date{Received   /   Accepted     }

\abstract
{}
{This paper is a continuation of an ongoing study of the evolutionary 
processes affecting cluster galaxies.}
{Both CCD R band and H$\alpha$ narrow-band imaging was used to
determine photometric parameters ($m_{r}$, $r_{24}$, H$\alpha$ flux,
and equivalent width) and derive star formation rates for 227 CGCG 
galaxies in 8
low-redshift clusters. The galaxy sample is a subset of CGCG galaxies
in an objective prism survey (OPS) of cluster galaxies for H$\alpha$
emission.} 
{It is found that detection of
emission-line galaxies in the OPS is 85\%, 70\%, and 50\% complete at
the mean surface brightness values of $1.25\times 10^{-19}$,
$5.19\times 10^{-20}$, and $1.76\times 10^{-20}$ W m$^{-2}$
arcsec$^{-2}$, respectively, measured within the R band isophote of 24
mag ${\rm arcsec}^{-2}$ for the galaxy.}
{The CCD data, together with
matched data from a recent H$\alpha$ galaxy survey of UGC galaxies
within $v \le 3000$ km ${\rm s}^{-1}$, will be used for
a comparative study of R band and H$\alpha$ surface photometry between
cluster and field spirals.}


\authorrunning{Thomas et al.}
\titlerunning{Star-forming galaxies in clusters}
\maketitle

\section{Introduction}
\label{intro}

While the transformation of cluster disc galaxies from predominantly
spiral to mainly lenticular galaxies over the past $\sim$ 5 Gyr is
well established (e.g. Butcher \& Oemler 1978, 1984; Dressler et
al. 1997; Fasano et al. 2000), the mechanism or mechanisms that have
affected this transformation are not so clear. However, a comparative
study of the rate, distribution, and morphological dependence of star
formation between cluster and field spirals appears to be a promising
enquiry that can help to disentangle some of the suggested
transformation processes. For example, ram-pressure stripping of the
cold interstellar gas of spirals by the hot ionised intracluster
medium (e.g. Gunn \& Gott 1972; Quilis, Moore \& Bower 2000) should be
most effective in the centres of rich clusters, and may lead to a
rapid truncation of the star forming disc, but provides no obvious
mechanism to promote circumnuclear star formation. On the other hand,
strangulation, i.e. the stripping of an hypothesised hot halo gas of
spirals (e.g. Larson et al. 1980; Bower \& Balogh 2004) should be a
more gradual process; simulations by Bekki et al. (2002) have shown
that this would lead to anemic spirals rather than truncation. Tidal
interactions with the cluster potential can induce star formation
across both bulge and disc (e.g. Byrd \& Valtonen 1990), whereas
low-velocity interactions between galaxies can be efficient at
triggering star formation in central regions (e.g. Kennicutt et
al. 1987; Mihos et al. 1992; Iono et al. 2004). In contrast, galaxy
harassment, i.e. frequent galaxy high-speed encounters within a 
cluster, are expected to trigger modest disc-wide response of star
formation for giant spirals (see Moore et al. 1999; Mihos 2004).

Systematic comparative studies of the massive star formation
properties of cluster and field galaxies have already produced
interesting results.  Moss \& Whittle (2000, 2005) undertook an
objective prism survey (OPS) of a complete magnitude-limited sample of
727 CGCG galaxies (Zwicky et al. 1960--68) in 8 low-redshift clusters.
These authors show an enhancement of circumnuclear starburst emission
in cluster spirals associated with a disturbed morphology that is
attributed to slow galaxy--galaxy encounters and major and minor
mergers (see also Moss 2006).  Koopmann \& Kenney (2004a,b) have pioneered
a comparative study of the massive star formation properties of Virgo
cluster and isolated bright ($M_{\rm B} < -18$) S0--Scd galaxies via
analyses of R and H$\alpha$ surface photometry.  They show that the
median total normalised massive star formation rate is reduced by a
factor of 2.5 for cluster galaxies as compared to the field. Few of
the cluster or isolated galaxies are anemic, suggesting that
strangulation is not a major contributory factor in the reduced star
formation rates of Virgo spirals; rather, this reduction is caused by
spatial truncation of the star forming discs. In addition, several of
the truncated galaxies show evidence of recent tidal interaction or
minor mergers, such as enhanced central star formation rates and
disturbed stellar discs.

It is intended to extend the analyses of Koopmann \& Kenney to the clusters
studied by Moss \& Whittle using R band and narrow-band H$\alpha$
imaging obtained for a 227 subset of CGCG galaxies of mainly types Sa
+ later in the OPS. Clusters in the OPS include those of greater
central galaxy density (most especially Abell 1367 and the Coma
cluster) where on-going environmental effects on galaxy morphology and
transformation may be expected to be even more pronounced than for the
Virgo cluster. In the present paper, we discuss observational data and
data reduction for imaging data, and present global photometric
properties and derived star formation rates. Completeness limits for
the OPS are also determined.  A second paper (Bretherton et al., in
preparation) will give results of a comparative study of R band and
H$\alpha$ surface photometry between the sample of 227 CGCG cluster
galaxies, and sets of galaxies, matched according to morphology and
absolute magnitude to the cluster sample, taken from the recent
H$\alpha$ galaxy survey of UGC galaxies within $v \le 3000$ km ${\rm
s}^{-1}$ (H$\alpha$GS, Shane 2002; James et al. 2004).

In the present paper, sample selection and observations of the cluster
data are discussed in section \ref{sselect}. The data reduction
procedures and photometry are outlined, and global parameters derived
for all sample galaxies, in sections \ref{datred} and \ref{glp}
respectively. Section \ref{complet} uses a complete sample of Sa--Sc
galaxies within the cluster data to investigate the completeness of
the objective prism survey (OPS, see Moss \& Whittle 2000, 2005) on
which present sample selection is based. Conclusions of this paper are
given in section \ref{concl}.

\section{Sample selection and observations}
\label{sselect}

\subsection{Selection of cluster sample}
\label{cfsample}

The galaxy samples chosen for the present study are sub-samples of
low-redshift cluster galaxies previously surveyed by Moss \& Whittle
(2000; 2005).  These authors undertook an extensive objective prism
survey of combined H$\alpha$ + [NII] emission for CGCG galaxies
(Zwicky et al. 1960--68) in 8 low-redshift Abell clusters (viz.  Abell
262, 347, 400, 426, 569, 779, 1367, and 1656). This objective prism
survey (hereafter, OPS) comprised a total of 727 CGCG galaxies,
including double galaxy components ($m_{p} \le 15.7$), and is
essentially complete within $r \le 1.5$ Abell radii of the cluster
centres. (The Abell radius, ${\rm r}_{\rm A}$ = 2.00 Mpc, assuming a
Hubble constant, $H_{0} = 75$ km ${\rm s}^{-1}$ ${\rm Mpc}^{-1}$, see
Abell 1958). For convenience, Table 1 summarises basic data for the
clusters surveyed, adapted from Moss \& Whittle (2000) and Moss
(2006).

\begin{table*}
\begin{center}
\begin{minipage}{130mm}
\caption{Surveyed clusters}
\label{clusters}
\begin{tabular}{lrrrrrrrrrrrr} \hline
Cluster & \multicolumn{6}{c}{Cluster centre} & \multicolumn{1}{c}{${\rm r}_{\rm A}$} &
\multicolumn{2}{c}{ ${\rm r}_{\rm vir}$ } & 
\multicolumn{1}{c}{ $z_{o}^{\dagger}$ } &
\multicolumn{1}{c}{ $\sigma_{v}^{\dagger}$ } & \multicolumn{1}{c}{$n^{\dagger}$} \\ \cline{2-7} \cline{9-10}
 & \multicolumn{4}{c}{R.A. (1950) Dec.} & \multicolumn{1}{c}{$l$} &
\multicolumn{1}{c}{$b$} & \multicolumn{1}{c}{(arcmin)} & 
\multicolumn{1}{c}{(Mpc)} & \multicolumn{1}{c}{(${\rm r}_{\rm A}$)} &&
\multicolumn{1}{c}{(km ${\rm s}^{-1}$)} & \\ \hline 
 &&&&&&&& \\
 Abell 262  & 
 1$^{\rm h}$ \hspace{-0.85em} &
 49\fm9 \hspace{-0.75em} &
 35$^{\circ}$ \hspace{-0.85em} &
 54$^{\prime}$ \hspace{-0.65em} &
 136\fdg59 \hspace{-0.50em} &
 -25\fdg09 \hspace{-0.50em} &      105 \hspace{0.70em} & 1.83 & 0.86 &     0.0167&    
537 \hspace{0.60em}   &    38 \\ 
 Abell 347  &    2& 22.7&    41& 39&   141.17&-17.63&       91 \hspace{0.70em} & 1.87 & 0.87 &     0.0192&    
550 \hspace{0.60em}  &    14 \\                                         
 Abell 400  &    2& 55.0&     5& 50&   170.25&-44.93&       72 \hspace{0.70em} & 1.32 & 0.62 &     0.0227&    
392 \hspace{0.60em}  &    10 \\  
 Abell 426  &    3& 15.3&    41& 20&   150.39&-13.38&       96 \hspace{0.70em} & 3.67 & 1.72 &     0.0177&   
1076 \hspace{0.60em} &   55 \\                                         
 Abell 569  &    7&  5.4&    48& 42&   168.58& 22.81&       88 \hspace{0.70em} & 1.42 & 0.66 &     0.0198&    
417 \hspace{0.60em}  &    26 \\                                         
 Abell 779  &    9& 16.8&    33& 59&   191.07& 44.41&       75 \hspace{0.70em} & 0.98 & 0.46 &     0.0232&    
290 \hspace{0.60em}  &    11 \\
 Abell 1367 &   11& 41.9&    20&  7&   234.81& 73.03&       80 \hspace{0.70em} & 2.58 & 1.21 &     0.0216&    
762 \hspace{0.60em}  &    62 \\                                         
 Abell 1656 &   12& 57.4&    28& 15&    58.09& 87.96&       74 \hspace{0.70em} & 3.01 & 1.41 &     0.0232&    
890 \hspace{0.60em}  &   131 \\ \hline
\end{tabular}
$^{\dagger}$ E--S0/a galaxies only

Cluster centres are taken from Abell, Corwin \& Olowin (1989). The
Abell radius (Abell 1958), ${\rm r}_{\rm A}$, corresponds to $\sim 1.5
h^{-1}$ Mpc. Values for the virial radius, ${\rm r}_{\rm vir}$, are
from Moss (2006), where $h=0.75$ is assumed. Cluster mean redshifts,
$z_{0}$, and velocity dispersions, $\sigma_{v}$, are based on $n$
redshifts of galaxies of types E--S0/a only. These objects are more likely to be in dynamic equilibrium with the cluster potential and follow a Gaussian distribution, as expected for virially relaxed galaxies within the cluster (see Moss 2006).  The mean
redshift has been corrected to the centroid of the Local Group
following RC2 (de Vaucouleurs, de Vaucouleurs \& Corwin 1976).

\end{minipage}
\end{center}
\end{table*}

The sub-samples of the OPS were restricted to galaxies with
velocities within $3\sigma$ of the cluster mean, and excluded known
AGNs since the present study is mainly concerned with star formation
properties.  Two principal sub-samples were chosen: (a) galaxies of
types Sa--Sc; this sample is complete for six clusters (viz. all
clusters of the OPS except Abell 262 and 347), and (b) emission-line
galaxies (ELGs) of the OPS of types S0/a and later (including
`peculiar' galaxies whose types are outside the Hubble sequence, see
Moss \& Whittle 2000).

\subsection{Observations}
\label{obs}

The observations of the cluster sample were taken on the 1.0m Jacobus 
Kapteyn Telescope (JKT) and the 2.6m Nordic Optical Telescope (NOT) both
situated on the island of La Palma.  The JKT data consist of 2 weeks of
observations taken in 1994 and 1998, some data taken in 1997, and a handful
of galaxies observed in earlier service time, although the majority of the 
service galaxies were also observed in subsequent runs.  Altogether, the
JKT runs provided H$\alpha$ and R band data for 143 sample galaxies, with
repeat measurements for 14 of these.  A 3 night observing run using
ALFOSC (Andalucia Faint Object Spectrograph and Camera) on the NOT in
January 2005 provided data for a further 87 galaxies, plus repeat
measurements for eight JKT galaxies and two NOT galaxies.

\begin{figure}
\includegraphics[width=87mm]{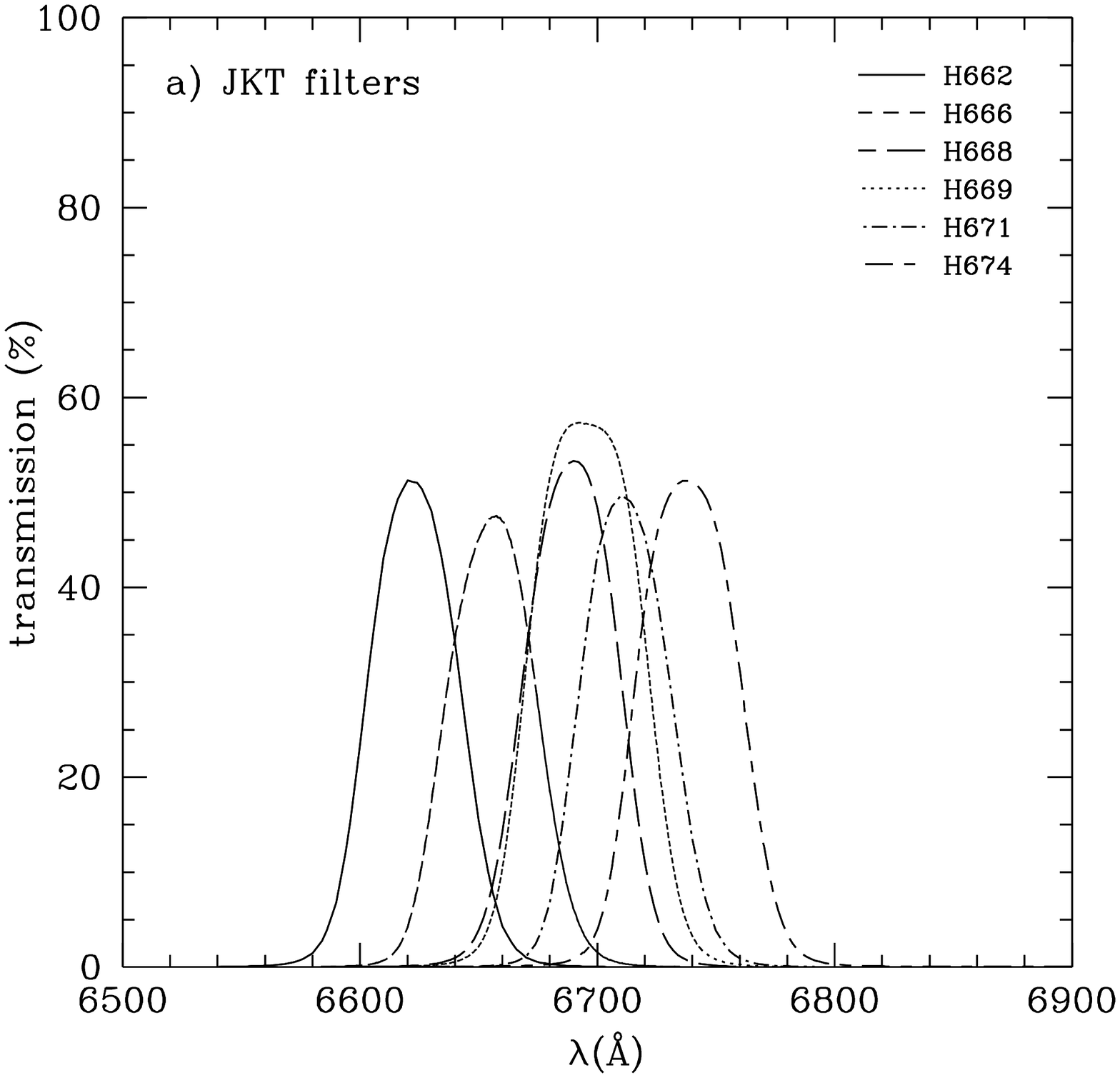}
\includegraphics[width=87mm]{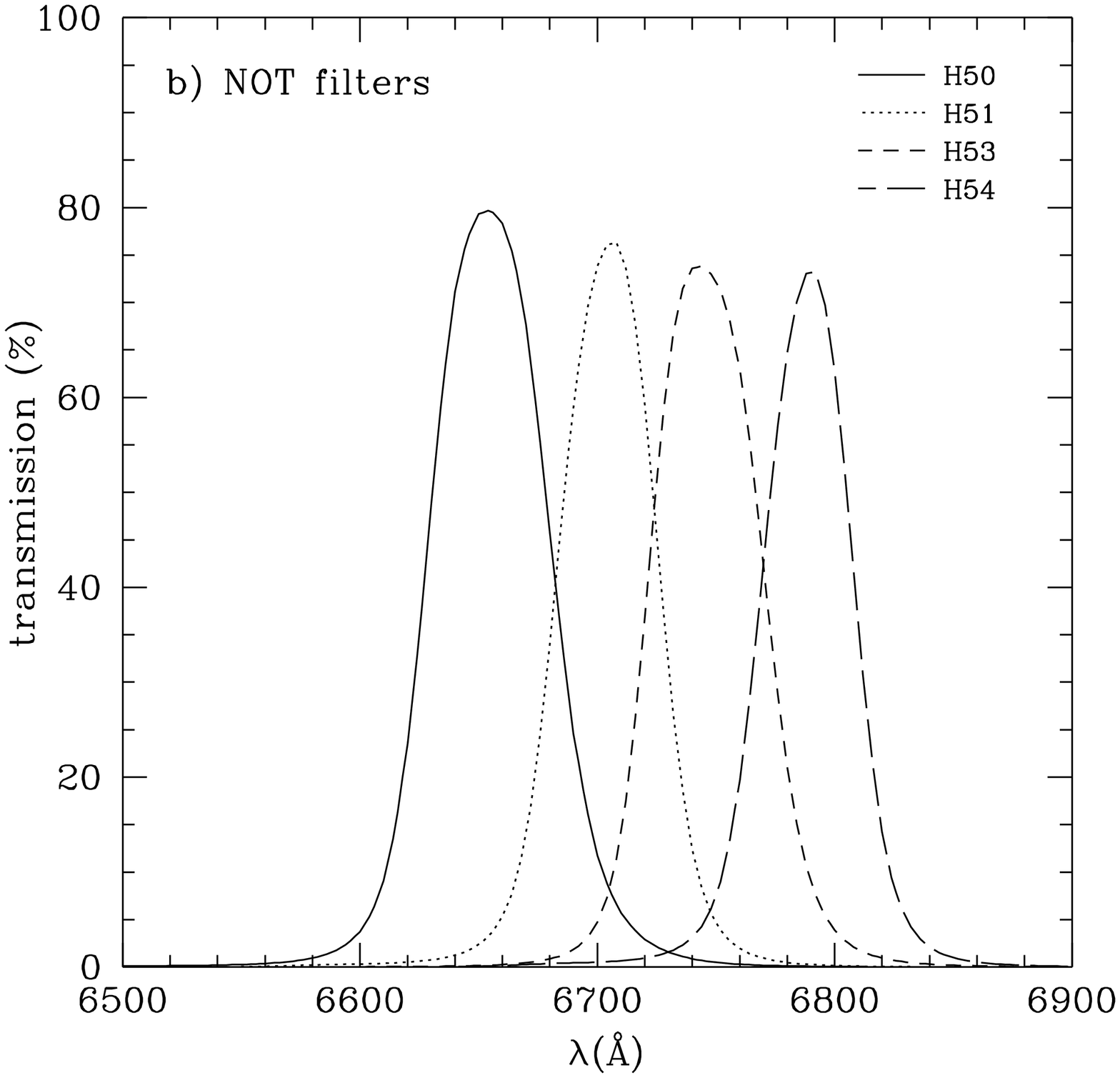}
\caption{Transmission profiles for both JKT and NOT redshifted narrowband 
H$\alpha$ filters. Filter H669 in the JKT plot was used only for service data.}
\label{fig:trans}
\end{figure}

The galaxies in the cluster sample have recession velocities in the range
$\sim 3000 - 9500$ km ${\rm s}^{-1}$. Therefore, in order to cover the
redshifted H$\alpha$ emission in all objects, it was necessary to use a 
series of narrowband filters.  These have a typical passband width of 40--55 
\AA, with peak wavelengths ranging from 6626--6788 \AA. Their transmission
profiles for the JKT and NOT observations are shown in Figures
\ref{fig:trans}a and \ref{fig:trans}b respectively.

The majority of the JKT observations made use of $1024 \times 1024$
pixel TEK4 CCD camera, with field of view $5.6 \times 5.6$ arcmin; for
the few galaxies surveyed in the early service observations, a $1204
\times 1124$ pixel EEV7 CCD camera was used, with field of view $6.2
\times 5.8$ arcmin.  R band continuum observations were taken using a
Harris R band filter, with central wavelength of 6373 \AA\/ and passband
width of 1491 \AA.

The more recent NOT observations utilised the standard ALFOSC CCD8 
$2048 \times 2048$ pixel back-illuminated device with a field of view
approximately $6.5 \times 6.5$ arcmin.  R band continuum observations 
were taken using a Bessell R filter with central wavelength of 6500 \AA\/ 
and passband width of 1300 \AA. The broad R band filter is used for continuum subtraction, rather than off-band H$\alpha$ filters, due to the much shorter exposure times required. James et al. (2004) found that scaled R band exposures gave excellent continuum subtraction when taken in dark sky conditions. The R band data are also used to trace the older, underlying stellar population of the sample galaxies in subsequent analysis.

As well as the galaxy images, a number of calibration frames were also 
taken.  These included bias frames taken at the start and end of each
night, sky flats (at least three observations of blank areas of sky, in
each filter used, taken at twilight every night), and photometric 
standards observed throughout the night in the R band filter to monitor
photometric conditions and calibrate the galaxy data.  Spectrophotometric
standards were also observed nightly through each filter.

\subsection{The final observed cluster sample}
\label{csample}

Combining data from all runs, a total of 230 CGCG galaxies were
observed.  These galaxies are listed in Table 2, where the first
three columns of the table give the CGCG number of the galaxy, 
the observing run, and narrow-band filter used to obtain the H$\alpha$
data respectively.

The observed galaxies include all Sa--Sc galaxies in six Abell
clusters (A400, 426, 569, 779, 1367, 1656) and emission-line galaxies
of types S0/a and later in the 8 clusters of the OPS.  Selection
criteria for the galaxies (see section \ref{cfsample} above) are a
velocity within $3\sigma$ of the cluster mean, a magnitude, $m_{p} \le
15.7$, a radial distance from the cluster centre, $r \le 1.5{\rm
r}_{\rm A}$, and that the galaxies are not known AGNs.  Some
18 galaxies are listed in Table 2 that do not strictly meet these
selection criteria: 13 galaxies have velocities greater than $3\sigma$
from the cluster mean (viz. CGCG 415-028, 415-031, 415-033, 415-035,
415-051, 415-58, 540-058, 540-071, 181-017, 181-023, 126-075, 127-002,
and 127-012B) and 5 galaxies are fainter than $m_{p} = 15.7$ (viz. CGCG
540-112B, 234-088B, 127-025B, 97-092A and 127-051B). These were observed in the preliminary phase of the project and have been included for completeness, but are not included in any of the statistical analyses in the present or subsequent papers. Twenty-four more galaxies (mainly in Abell 1367) lie beyond 1.5 Abell radii from the
cluster centre.

\section{Data reduction}
\label{datred}

\subsection{Continuum subtraction}
\label{csubtr}

Standard IRAF and {\it Starlink} tasks were used to debias, flat-field and
remove cosmic ray defects from all images, as well as to align pairs
of R band and H$\alpha$ images.  Some galaxies in the cluster sample
had uneven sky backgrounds after processing.  For the NOT data this
was simply a gradient across the H$\alpha$ images.  The JKT data were
more variable, with some images having smooth, even backgrounds,
whilst others showed gradients, marks or rings.  Background
variability was reduced by fitting a surface to the sky in affected
images.  All stars and galaxies in a frame were identified and masked
out, and a polynomial fit was made to the remaining background using
the {\it Starlink KAPPA} command {\it surfit}.  This was subtracted
from the original image, removing much of the variability and
typically reducing sky background errors in affected frames by a
factor of 15.

A critical stage in the data reduction process is the scaling and
subtraction of the continuum contribution from the narrowband
H$\alpha$ images.  Scaling factors were calculated from a comparison
of 9 field stars in each galaxy image pair and from standard star
observations.  A mean scaling factor was then found for each filter
pair across the full dataset. Scaling factors were consistent for all
objects observed in the same H$\alpha$ filter during photometric
conditions.

The scaled continuum images were subtracted from the H$\alpha$ images to
leave continuum-free H$\alpha$ emission.  This latter includes a contribution
from the neighbouring [NII] lines ($\lambda\lambda$ 6548, 6585\AA) that lie
within the filter bandpass.

\subsection{Calibration}
\label{calib}

Images were calibrated by using R band standard star observations to 
determine extinction corrections and zeropoints for each night.
Six galaxies (viz. CGCG 415-033, 522-011, 522-021, 522-053, 522-072, and
539-031) were observed in non-photometric 
conditions on the second night of the NOT run.  For these galaxies separate
R band calibration exposures were taken under photometric conditions on
the following night and used to calibrate the non-photometric data.
  
Calibrated images were normalised to 1 count = 25 mag. The apparent
magnitude of each galaxy in a given filter, $m_{filt}$, is then simply
given by,

$m_{filt} = 25 - \log k_{g} - A_{\rm R}$

where $k_{g}$ is the number of sky-subtracted counts measured directly from
the R band or continuum subtracted H$\alpha$ images, and ${\rm A}_{\rm R}$
is a correction for Galactic extinction.  Values of ${\rm A}_{\rm R}$ were
based on the Galactic extinction maps of Schlegel et al. (1998) converted to
R band extinction using the methods of Cardelli et al. (1989).

For narrow-band images, the measured magnitude, $m_{H\alpha}$, of a galaxy
can be considered as an equivalent magnitude in the R band. Following the
calibration of R(Cousins) from Bessell (1979), $m_{H\alpha} = 0$
corresponds to a flux density of $2.24 \times 10^{-12}$~W~$m^{-2}$ 
${\rm \AA^{-1}}$; this relation was used to determine H$\alpha$ flux
values.  Finally these flux values were corrected both for the centering
of the H$\alpha$ emission in the narrow-band filter, and for slight 
over-subtraction of the continuum due to the wavelength overlap of the 
R and narrow-band filters, to obtain final H$\alpha$ flux values used for
the subsequent analysis. Values of the H$\alpha$ equivalent width (EW) for individual galaxies
were obtained by dividing the H$\alpha$ flux by the corresponding R band
flux density.  

\subsection{Photometry}
\label{phot}

Photometry of individual galaxies was performed using a series of
apertures of increasing radius/semi-major axis for each galaxy.
Typically between 25 and 60 apertures were used to cover the full
galaxy and measure out into the sky background.  Elliptical apertures
were generally used with major to minor axis ratio taken from NED or
from Moss \& Whittle (unpublished), and with position angles
determined from the images themselves using the {\it Starlink}
function {\it Object Detection} in {\it GAIA}.  For some objects with
no published axis ratio values, or for which the given axis ratio did
not appear to fit well the outer isophotes of the galaxy, axis ratios
were estimated from the data.  For face-on galaxies, or those with
peculiar shapes, circular apertures were employed.

The centre of each galaxy was generally found by centroiding on
the nucleus of the R band image in {\it GAIA}.  In a few cases it was found
necessary to determine the centre by eye.  The R band was preferred for the
centroiding process due to the relative smoothness of the R band emission
compared to the clumpier H$\alpha$ data.

Before photometry was carried out, all stars or stellar residuals were 
masked from each image.  Counts were then measured in each aperture, with
sky subtraction performed using an annular sky region outside the largest
galaxy aperture.  These were converted to H$\alpha$ flux or R band flux
density according to the procedure given in section \ref{calib}, and were then
combined to create growth curves for R band flux density, H$\alpha$ flux
and EW (H$\alpha$ flux/R flux density), which form the basis of the 
remainder of this study.

\section{Global parameters}
\label{glp}

Global parameters for the observed galaxies are listed in Table 2.
These parameters include the absolute B magnitude; the isophotal
radius, $r_{24}$, which is the semi-major axis of the isophote equal
to 24 mag ${\rm arcsec}^{-2}$; the apparent magnitude, $m_{R}$, which
corresponds to the light enclosed within this isophote; values of
H$\alpha$ flux and EW; and the total star formation rate for the
galaxy.  Each of these parameters is explained in more detail the
sections that follow.

\begin{table*}
\caption{Global parameters for 227 CGCG galaxies in eight low redshift Abell clusters}
\label{gparams}

\begin{tabular}{@{}l@{\hspace{0.9em}}l@{\hspace{0.9em}}l@{\hspace{0.9em}}r@{\hspace{0.9em}}
l@{\hspace{0.9em}}c@{\hspace{0.9em}}c@{}c@{}c@{}c@{}c@{}
c@{}c@{}c@{}}

\hline
 CGCG & Run & H$\alpha$ & $M_{\rm B}$ & Type & \multicolumn{2}{c}{R.A. (2000) Dec.} & 
\multicolumn{1}{c}{a/b} & \multicolumn{1}{c}{PA} & 
\multicolumn{1}{c}{$r_{24}$} & \multicolumn{1}{c}{$m_{\rm R}$} & 
\multicolumn{1}{c}{H$\alpha$ flux} & \multicolumn{1}{c}{H$\alpha$ EW} & 
\multicolumn{1}{c}{SFR} \\
 && filter &&& (h \hspace{0.5em} m \hspace{0.5em} s) & 
(\degr \hspace{0.5em} \arcmin \hspace{0.5em} \arcsec) && \multicolumn{1}{c}{($^{\circ}$)} & \multicolumn{1}{c}{(\arcsec)} &&
\multicolumn{1}{@{}c@{}}{($10^{-17} {\rm W} {\rm m}^{-2}$)} & \multicolumn{1}{c}{(nm)} & 
\multicolumn{1}{@{}c}{(M$_{\odot}$yr$^{-1}$)} \\ \hline
\multicolumn{2}{@{}l}{{\it Abell 262}} &&&&&&&&&&&& \\
521-073  & JKT94 & H668 & -20.9 & S pec                   & 01 44 35.5  & +37 41 45 & 1.6 &  17 & 31 & 13.17 &  58(5) &  4.6(0.4) &  4.8 \\
521-074  & JKT94 & H668 & -19.8 & S: pec                  & 01 44 38.4  & +34 39 37 & 4.5 &  86 & 20 & 14.70 &  12(1) &  3.9(0.4) &  1.0 \\
521-078  & JKT94 & H668 & -19.6 & Sc/SBc                  & 01 45 51.8  & +35 06 39 & 1.4 & 144 & 18 & 14.27 &  19(2) &  4.2(0.4) &  1.6 \\
522-003  & JKT94 & H666 & -19.3 & pec                     & 01 46 56.8  & +34 46 21 & 1.5 &  87 & 17 & 13.97 &  16(2) &  2.7(0.3) &  1.4 \\
522-006  & JKT94 & H668 & -19.3 & SAbc: pec               & 01 47 43.6  & +35 01 21 & 1.0 & \ldots & 17 & 13.91 &  17(2) &  2.6(0.3) &  1.4 \\
522-011  & NOT   & H50  & -19.0 & S0/a                    & 01 49 18.1  & +34 58 06 & 1.7 &  40 & 16 & 14.70 &  15(1) &  5.1(0.4) &  1.2 \\
522-020  & JKT94 & H666 & -21.6 & SBb                     & 01 50 44.2  & +35 17 04 & 1.7 & 141 & 68 & 12.32 &  45(7) &  1.6(0.3) &  3.8 \\
522-021  & NOT   & H50  & -20.5 & S                       & 01 50 47.9  & +35 55 58 & 3.8 &  60 & 72 & 12.92 &   3(3) &  0.2(0.2) &  0.2 \\
522-024  & JKT94 & H668 & -20.3 & SA: pec                 & 01 50 47.9  & +35 55 58 & 1.5 & 163 & 34 & 13.16 &  36(4) &  2.9(0.3) &  3.0 \\
522-025  & JKT94 & H671 & -19.2 & SAbc:                   & 01 52 02.8  & +36 07 52 & 1.4 &   6 & 19 & 14.04 &  20(2) &  3.5(0.3) &  1.7 \\
522-038  & NOT   & H51  & -20.7 & Sc/SBc                  & 01 52 45.9  & +36 37 10 & 1.0 & \ldots & 37 & 12.72 &  45(5) &  2.5(0.3) &  3.7 \\
522-041  & NOT   & H51  & -20.3 & SABc                    & 01 52 53.9  & +36 03 10 & 1.0 & \ldots & 32 & 13.25 &  59(4) &  5.3(0.4) &  4.9 \\
522-053  & NOT   & H51  & -19.5 & S0:                     & 01 53 57.0  & +36 38 28 & 2.3 &  35 & 28 & 13.97 &   5(1) &  0.9(0.2) &  0.4 \\
522-055  & JKT94 & H668 & -20.6 & SBc                     & 01 54 19.8  & +36 37 47 & 4.2 & 138 & 36 & 13.10 &   8(3) &  0.6(0.2) &  0.6 \\
522-058  & JKT92 & H669 & -20.5 & SBa                     & 01 54 53.8  & +36 55 05 & 1.2 &  73 & 38 & 13.05 &  20(4) &  1.4(0.3) &  1.7 \\
522-069  & JKT94 & H668 & -19.7 & SAc:                    & 01 55 58.5  & +37 07 46 & 1.1 &  36 & 20 & 13.31 &  14(3) &  1.2(0.3) &  1.1 \\
522-072  & NOT   & H50  & -19.2 & E/S0:: pec              & 01 56 21.4  & +35 34 21 & 1.2 & 126 & 15 & 14.22 &   7(1) &  1.5(0.3) &  0.6 \\
522-074  & JKT94 & H668 & -19.9 & Sc                      & 01 56 21.1  & +37 27 08 & 8.7 & 123 & 24 & 15.78 &   2(1) &  2.1(0.9) &  0.2 \\
522-077  & JKT94 & H668 & -19.1 & SBb: pec                & 01 56 30.5  & +37 20 08 & 1.2 &  43 & 15 & 14.02 &   8(2) &  1.4(0.3) &  0.7 \\
522-079  & JKT94 & H668 & -19.4 & SA:c:                   & 01 56 39.9  & +35 35 32 & 1.8 & 135 & 17 & 14.12 &  13(2) &  2.4(0.3) &  1.1 \\
522-081  & JKT92 & H669 & -20.3 & S                       & 01 56 46.1  & +36 53 11 & 2.0 &  54 & 46 & 13.64 & \ldots & \ldots & \ldots \\
522-086  & JKT94 & H668 & -22.2 & SAB:c                   & 01 57 42.2  & +35 54 58 & 1.5 & 126 & 71 & 11.59 &181(18) &  3.4(0.3) & 15.1 \\
522-096  & JKT94 & H668 & -20.3 & Sc                      & 01 59 06.7  & +36 03 47 & 8.0 & 107 & 31 & 14.95 &   4(1) &  1.4(0.3) &  0.3 \\
522-100  & JKT94 & H666 & -19.7 & SB(s)dm                 & 02 00 11.2  & +37 36 12 & 1.2 &   6 & 30 & 13.99 &  15(2) &  2.5(0.3) &  1.2 \\
522-102  & JKT94 & H666 & -21.1 & SB:ab                   & 02 00 54.9  & +38 12 39 & 2.6 &  83 & 60 & 13.00 &  31(4) &  2.1(0.3) &  2.6 \\
\multicolumn{2}{@{}l}{{\it Abell 347}} &&&&&&&&&&&& \\
538-037  & JKT94 & H668 & -20.2 & Sc                      & 02 15 54.1  & +42 49 26 & 4.0 & 105 & 29 & 15.47 &   4(1) &  2.6(0.7) &  0.4 \\
538-043  & JKT92 & H669 & -20.1 & pec                     & 02 20 03.8  & +41 16 28 & 1.5 & 120 & 33 & 14.00 &  42(3) &  7.2(0.5) &  4.7 \\
538-046  & JKT94 & H668 & -19.7 & SA:b:                   & 02 20 35.0  & +41 34 27 & 1.2 &  64 & 22 & 14.17 &   8(4) &  1.6(0.7) &  0.9 \\
538-048  & JKT92 & H669 & -20.5 & S pec                   & 02 21 23.3  & +42 52 35 & 3.6 & 126 & 51 & 13.95 &  18(2) &  2.9(0.3) &  2.0 \\
523-028  & JKT98 & H671 & -21.5 & SA(s)b pec$^{\dagger}$  & 02 21 28.7  & +39 22 32 & 1.6 &  51 & 65 & 12.28 & \ldots & \ldots & \ldots \\
523-029  & JKT98 & H671 & -20.2 & SB(s)a pec$^{\dagger}$  & 02 21 32.6  & +39 21 25 & 3.8 &  89 & 46 & 14.43 &   9(3) &  2.2(0.7) &  1.0 \\
538-054  & JKT94 & H671 & -19.5 & Sa:                     & 02 22 50.4  & +42 09 29 & 2.0 &  93 & 22 & 14.75 &   5(1) &  1.6(0.3) &  0.5 \\
538-063  & JKT94 & H668 & -19.8 & Sbc                     & 02 24 47.3  & +42 01 28 & 2.7 &  48 & 26 & 14.37 &   8(1) &  1.8(0.3) &  0.9 \\
539-023  & JKT94 & H668 & -20.7 & SAc                     & 02 26 05.2  & +42 08 39 & 1.4 &  63 & 18 & 13.14 &   1(3) &  0.1(0.2) &  0.1 \\
539-024  & JKT92 & H669 & -20.5 & SBb                     & 02 26 46.3  & +41 50 04 & 2.0 & 142 & 50 & 13.24 &  35(4) &  3.0(0.3) &  4.0 \\
539-025  & JKT94 & H666 & -19.8 & SB pec                  & 02 26 53.1  & +41 41 25 & 1.2 &  11 & 28 & 13.48 &  13(2) &  1.4(0.3) &  1.4 \\
539-029  & JKT94 & H671 & -19.7 & S                       & 02 27 31.2  & +41 55 51 & 2.0 &  47 & 30 & 13.69 &  14(2) &  1.7(0.3) &  1.5 \\
539-030  & NOT   & H51  & -20.8 & Sb:                     & 02 27 34.6  & +41 58 39 & 1.6 & 105 & 35 & 13.06 &  52(5) &  3.9(0.4) &  5.8 \\
539-031  & NOT   & H51  & -19.9 & S0/a                    & 02 27 36.7  & +42 00 28 & 1.3 & 142 & 20 & 13.99 &  25(2) &  4.4(0.4) &  2.8 \\
539-036  & JKT92 & H669 & -20.8 & Sab                     & 02 31 14.3  & +40 23 25 & 3.3 & 117 & 51 & 13.87 &  15(2) &  2.2(0.3) &  1.6 \\
539-038  & JKT92 & H669 & -19.3 & S pec                   & 02 31 26.6  & +40 14 51 & 1.0 & \ldots & 23 & 14.24 &  14(2) &  3.0(0.3) &  1.6 \\
539-044  & JKT94 & H668 & -20.1 & pec                     & 02 33 40.2  & +41 20 47 & 3.0 & 143 & 28 & 14.17 &  21(2) &  4.2(0.4) &  2.4 \\
\multicolumn{2}{@{}l}{{\it Abell 400}} &&&&&&&&&&&& \\
415-020  & NOT   & H53  & -21.3 & S0                      & 02 53 41.2  & +06 15 55 & 3.8 & 142 & 80 & 12.55 &  30(6) &  1.4(0.3) &  5.3 \\
415-021  & JKT94 & H674 & -20.9 & SAB:c                   & 02 53 58.8  & +05 59 16 & 1.1 &   7 & 29 & 13.83 &  33(3) &  4.8(0.4) &  5.8 \\
415-025  & JKT94 & H674 & -20.5 & S                       & 02 55 19.9  & +06 07 29 & 1.0 &   0 & 20 & 13.80 &  33(3) &  4.6(0.4) &  5.8 \\
415-028  & NOT   & H53  & -21.1 & SABc                    & 02 55 46.9  & +06 13 03 & 1.0 & \ldots & 36 & 13.24 &  27(3) &  2.4(0.3) &  4.8 \\
415-030  & JKT94 & H674 & -21.5 & Sc                      & 02 55 57.6  & +06 29 41 & 2.5 & 165 & 43 & 13.70 &  26(3) &  3.3(0.3) &  4.6 \\
415-031  & NOT   & H53  & -21.0 & Sc                      & 02 56 17.6  & +04 31 46 & 3.5 & 158 & 44 & 14.67 &   8(1) &  2.5(0.3) &  1.3 \\
415-032  & NOT   & H53  & -21.0 & SBbc                    & 02 56 22.7  & +06 09 19 & 2.5 &   6 & 32 & 14.63 &  11(1) &  3.6(0.3) &  2.0 \\
415-033  & NOT   & H53  & -20.5 & S0:: pec                & 02 56 28.3  & +04 36 38 & 1.0 & \ldots & 19 & 14.02 &  16(2) &  2.9(0.3) &  2.8 \\
415-035  & NOT   & H53  & -21.4 & SBa                     & 02 56 43.1  & +07 20 01 & 1.0 & \ldots & 39 & 12.61 &  16(5) &  0.8(0.2) &  2.9 \\
415-037  & NOT   & H53  & -21.3 & Sc                      & 02 56 46.9  & +04 58 40 & 3.7 & 141 & 48 & 13.71 &   8(2) &  1.2(0.3) &  1.5 \\
415-038  & NOT   & H51  & -20.5 & S0$^{\dagger}$          & 02 56 56.6  & +06 12 20 & 1.3 &  40 & 19 & 14.20 &  13(1) &  2.9(0.3) &  2.4 \\
415-039  & NOT   & H53  & -21.0 & SA:b                    & 02 57 09.2  & +05 19 15 & 1.5 &  67 & 32 & 13.48 &  10(2) &  1.1(0.3) &  1.8 \\
415-048  & JKT94 & H671 & -21.2 & S                       & 02 58 29.7  & +06 18 23 & 1.7 &  47 & 34 & 13.18 &  36(4) &  2.9(0.3) &  6.4 \\
415-051  & NOT   & H53  & -21.0 & Sab$^{\dagger}$         & 02 59 52.3  & +06 31 59 & 1.8 & 156 & 32 & 13.68 &  15(2) &  2.0(0.3) &  2.7 \\
415-052  & NOT   & H51  & -20.4 & SA0/a:                  & 03 00 08.6  & +05 48 16 & 1.0 & \ldots & 23 & 13.70 &   1(2) &  0.1(0.2) &  0.1 \\
415-058  & NOT   & H53  & -20.6 & Sbc:                    & 03 04 39.2  & +05 26 35 & 1.8 &  57 & 29 & 14.20 &   8(1) &  1.7(0.3) &  1.4 \\
\end{tabular}
\end{table*}

\addtocounter{table}{-1}
\begin{table*}
\caption{-- continued}


\begin{tabular}{@{}l@{\hspace{0.9em}}l@{\hspace{0.9em}}l@{\hspace{0.9em}}r@{\hspace{0.9em}}
l@{\hspace{0.9em}}c@{\hspace{0.9em}}c@{}c@{}c@{}c@{}c@{}
c@{}c@{}c@{}}

\hline
 CGCG & Run & H$\alpha$ & $M_{\rm B}$ & Type & \multicolumn{2}{c}{R.A. (2000) Dec.} & 
\multicolumn{1}{c}{a/b} & \multicolumn{1}{c}{PA} & 
\multicolumn{1}{c}{$r_{24}$} & \multicolumn{1}{c}{$m_{\rm R}$} & 
\multicolumn{1}{c}{H$\alpha$ flux} & \multicolumn{1}{c}{H$\alpha$ EW} & 
\multicolumn{1}{c}{SFR} \\
 && filter &&& (h \hspace{0.5em} m \hspace{0.5em} s) & 
(\degr \hspace{0.5em} \arcmin \hspace{0.5em} \arcsec) &&
\multicolumn{1}{c}{($^{\circ}$)} & \multicolumn{1}{c}{(\arcsec)} &&
\multicolumn{1}{@{}c@{}}{($10^{-17} {\rm W} {\rm m}^{-2}$)} & \multicolumn{1}{c}{(nm)} & 
\multicolumn{1}{@{}c}{(M$_{\odot}$yr$^{-1}$)} \\ \hline
\multicolumn{2}{@{}l}{{\it Abell 426}} &&&&&&&&&&&& \\
540-047  & NOT   & H51  & -21.1 & Sb                      & 03 09 42.7  & +40 58 27 & 5.5 & 130 & 54 & 13.29 & \ldots & \ldots & \ldots \\
540-049  & JKT94 & H662 & -21.3 & S-Irr                   & 03 10 12.9  & +40 45 56 & 2.0 &  71 & 29 & 12.54 &  50(7) &  2.2(0.3) &  5.0 \\
540-058  & NOT   & H53  & -20.3 & Sb pec                  & 03 13 10.2  & +42 59 49 & 1.7 &  25 & 22 & 14.18 &  13(1) &  2.8(0.3) &  1.3 \\
525-009  & JKT94 & H666 & -21.0 & SBc                     & 03 14 40.5  & +39 37 03 & 1.4 & 139 & 25 & 12.93 &  55(5) &  3.5(0.3) &  5.5 \\
525-011  & NOT   & H50  & -20.9 & Sb                      & 03 14 58.0  & +39 20 56 & 2.2 &  37 & 40 & 13.61 &  21(2) &  2.6(0.3) &  2.1 \\
540-064  & JKT92 & H669 & -21.7 & SBb                     & 03 15 01.4  & +42 02 09 & 1.1 &  70 & 50 & 12.54 & 88(18) &  3.9(0.8) &  8.8 \\
540-067  & JKT94 & H668 & -20.1 & SA:a:                   & 03 15 20.5  & +41 36 45 & 1.3 &  37 & 18 & 14.11 &  17(2) &  3.2(0.3) &  1.7 \\
540-069  & JKT94 & H666 & -21.7 & SABc                    & 03 16 00.8  & +40 53 08 & 2.8 &   3 & 41 & 13.90 &  10(7) &  1.5(1.1) &  1.0 \\
540-070  & JKT94 & H668 & -21.1 & Sab                     & 03 16 01.0  & +42 04 28 & 2.4 & 165 & 46 & 13.15 &   5(4) &  0.4(0.2) &  0.5 \\
540-071  & NOT   & H53  & -20.3 & SA:a:                   & 03 16 02.9  & +42 55 18 & 1.4 &  77 & 27 & 13.95 &  16(2) &  2.7(0.3) &  1.6 \\
540-073  & NOT   & H50  & -21.6 & Sa                      & 03 16 26.1  & +41 31 50 & 5.7 &  67 & 82 & 12.95 &   2(3) &  0.1(0.2) &  0.2 \\
540-078  & NOT   & H51  & -20.7 & Sa:                     & 03 16 59.7  & +41 21 24 & 4.7 &  71 & 54 & 13.38 &   2(2) &  0.2(0.2) &  0.2 \\
540-083  & NOT   & H50  & -21.0 & Sab                     & 03 17 50.4  & +41 58 03 & 5.0 &  89 & 70 & 13.67 &   1(2) &  0.1(0.2) &  0.1 \\
540-084  & JKT92 & H669 & -21.5 & SBb                     & 03 17 52.3  & +43 18 15 & 1.8 &  68 & 57 & 12.70 & 35(16) &  1.8(0.8) &  3.5 \\
540-091  & JKT92 & H669 & -22.0 & SBc                     & 03 18 45.3  & +43 14 27 & 2.2 &   0 & 54 & 13.66 &  52(4) &  6.5(0.5) &  5.2 \\
540-093  & NOT   & H50  & -21.3 & SAb                     & 03 18 45.2  & +41 29 19 & 1.7 & 130 & 39 & 13.11 & \ldots & \ldots & \ldots \\
540-094  & JKT94 & H671 & -20.9 & Sbc                     & 03 18 53.4  & +40 35 45 & 2.4 &  62 & 27 & 13.45 &  20(3) &  2.1(0.3) &  2.0 \\
540-100  & JKT94 & H674 & -20.7 & Sc? pec                 & 03 19 27.4  & +41 38 07 & 2.5 & 119 & 37 & 13.58 &  12(2) &  1.4(0.3) &  1.2 \\
540-103  & JKT94 & H668 & -22.9 & pec:                    & 03 19 48.2  & +41 30 42 & 1.3 & 104 & ** & 10.44 &658(57) &  4.2(0.4) & 66.0 \\
540-106  & NOT   & H50  & -20.3 & Sa?                     & 03 20 05.2  & +40 54 22 & 3.0 & 178 & 38 & 13.81 & \ldots & \ldots & \ldots \\
540-112A & NOT   & H50  & -20.6 & ... pec                 & 03 21 19.9  & +41 55 55 & 1.0 & \ldots & 40 & 12.92 &  17(4) &  1.1(0.3) &  1.7 \\
540-112B & NOT   & H50  & -20.7 & S:... pec               & 03 21 20.0  & +41 55 44 & 5.0 & 112 & 59 & 13.84 &   3(2) &  0.5(0.2) &  0.3 \\
540-114  & NOT   & H50  & -20.6 & S:a:                    & 03 21 32.8  & +40 24 37 & 3.5 &  14 & 47 & 13.75 &   1(2) &  0.2(0.2) &  0.1 \\
540-121  & JKT92 & H669 & -21.4 & SB:b                    & 03 22 53.8  & +42 33 12 & 3.8 & 129 & 77 & 12.76 &  19(9) &  1.0(0.5) &  1.9 \\
541-005  & NOT   & H50  & -21.6 & Sb                      & 03 25 52.3  & +40 44 56 & 4.7 & 128 & 67 & 13.26 &  18(3) &  1.6(0.3) &  1.8 \\
541-006  & NOT   & H51  & -20.7 & SBb                     & 03 25 59.2  & +40 47 21 & 1.0 & \ldots & 43 & 12.45 &  29(6) &  1.2(0.3) &  2.9 \\
541-008  & NOT   & H51  & -22.0 & Sab                     & 03 26 27.6  & +40 30 29 & 5.0 &  68 & 88 & 12.73 &   6(4) &  0.3(0.2) &  0.6 \\
541-009  & JKT94 & H666 & -20.5 & SBc                     & 03 27 39.8  & +40 53 49 & 1.3 & 114 & 30 & 13.55 &  11(2) &  1.3(0.3) &  1.1 \\
541-011  & NOT   & H50  & -21.2 & SB:b: pec               & 03 28 27.8  & +40 09 16 & 2.7 &  76 & 44 & 13.26 &  74(5) &  6.8(0.5) &  7.5 \\
541-017  & JKT94 & H666 & -21.6 & pec:                    & 03 30 01.8  & +41 49 55 & 2.1 & 118 & 45 & 12.45 & 117(9) &  4.8(0.4) & 11.7 \\
\multicolumn{2}{@{}l}{{\it Abell 569}} &&&&&&&&&&&& \\
234-043  & JKT94 & H668 & -21.1 & SB:ab                   & 07 03 02.8  & +49 25 28 & 2.3 &  43 & 40 & 13.21 &  22(3) &  1.8(0.3) &  2.6 \\
234-050  & NOT   & H51  & -20.8 & SBa:                    & 07 06 00.3  & +50 30 37 & 2.8 & 131 & 48 & 12.76 &   3(4) &  0.2(0.2) &  0.4 \\
234-051  & NOT   & H51  & -20.5 & SBa                     & 07 05 59.0  & +50 45 22 & 1.3 & 124 & 41 & 13.00 &   8(3) &  0.6(0.2) &  0.9 \\
234-056  & JKT97 & H669 & -20.1 & S pec                   & 07 07 32.4  & +48 54 00 & 1.5 & 121 & 18 & 14.42 &  28(2) &  7.0(0.5) &  3.4 \\
234-057  & NOT   & H51  & -19.1 & pec                     & 07 07 51.4  & +48 24 55 & 1.0 & \ldots & 13 & 15.01 &   6(1) &  2.7(0.3) &  0.7 \\
234-060  & NOT   & H51  & -21.1 & SBb                     & 07 08 11.0  & +50 40 55 & 1.1 &  99 & 65 & 11.88 & 29(10) &  0.8(0.2) &  3.5 \\
234-061  & JKT94 & H671 & -19.5 & SAa:                    & 07 08 09.7  & +48 55 46 & 1.1 &  93 & 17 & 13.86 &   3(2) &  0.5(0.2) &  0.4 \\
234-062  & NOT   & H51  & -20.1 & SB:a:                   & 07 08 12.2  & +49 08 17 & 1.4 & 100 & 28 & 13.73 &   3(2) &  0.4(0.2) &  0.3 \\
234-065  & NOT   & H51  & -19.3 & SB: pec                 & 07 08 31.2  & +48 08 13 & 1.0 & \ldots & 16 & 14.67 &  11(1) &  3.5(0.3) &  1.3 \\
234-066  & JKT97 & H669 & -20.0 & pec:                    & 07 08 34.2  & +50 37 53 & 1.7 &   3 & 32 & 14.15 &  24(2) &  4.6(0.4) &  2.8 \\
234-067  & JKT94 & H671 & -20.0 & Sa:                     & 07 08 47.3  & +48 59 41 & 1.4 &  71 & 23 & 14.28 &  17(2) &  3.7(0.3) &  2.0 \\
234-069  & JKT97 & H669 & -19.5 & Sa:                     & 07 09 03.6  & +48 34 24 & 1.2 & 139 & 19 & 14.54 &  14(1) &  3.8(0.4) &  1.6 \\
234-071  & JKT94 & H666 & -19.7 & SB: pec                 & 07 09 15.9  & +49 49 13 & 1.2 & 118 & 22 & 14.00 &  24(2) &  4.1(0.4) &  2.9 \\
234-077  & JKT94 & H671 & -19.9 & S0:                     & 07 09 39.4  & +47 54 54 & 1.7 & 140 & 29 & 13.67 & \ldots & \ldots & \ldots \\
234-079A & JKT94 & H671 & -19.8 & S: pec                  & 07 09 54.8  & +47 54 28 & 1.6 & 120 & 22 & 14.44 &  14(1) &  3.5(0.3) &  1.7 \\
234-079B & JKT94 & H671 & -19.9 & S: pec                  & 07 09 53.6  & +47 54 47 & 2.7 &  16 & 35 & 14.31 &  10(1) &  2.3(0.3) &  1.2 \\
234-085  & NOT   & H51  & -19.3 & \ldots                  & 07 10 52.2  & +48 12 13 & 1.4 & 133 & 15 & 15.01 &   6(1) &  2.7(0.3) &  0.7 \\
234-088A & NOT   & H51  & -20.4 & Sab                     & 07 11 01.2  & +48 30 47 & 3.0 & 160 & 48 & 13.32 &   9(3) &  0.8(0.2) &  1.0 \\
234-088B & NOT   & H51  & -18.8 & \ldots                  & 07 11 00.8  & +48 31 56 & 1.7 & 149 & 16 & 14.95 &   3(1) &  1.4(0.3) &  0.4 \\
234-090  & JKT94 & H668 & -20.5 & Sbc                     & 07 10 58.3  & +49 01 22 & 4.0 & 103 & 33 & 14.83 &   5(1) &  1.8(0.3) &  0.6 \\
234-092  & NOT   & H51  & -19.4 & Sa:                     & 07 11 08.8  & +49 53 58 & 1.0 & \ldots & 21 & 14.26 &  10(1) &  2.2(0.3) &  1.2 \\
234-093  & JKT94 & H668 & -20.8 & SBb                     & 07 11 28.0  & +48 14 24 & 1.4 &  54 & 41 & 12.88 &  18(4) &  1.1(0.3) &  2.2 \\
234-094  & JKT97 & H669 & -19.8 & S-Irr                   & 07 11 30.3  & +49 04 50 & 1.4 & 175 & 25 & 14.44 &  12(1) &  3.1(0.3) &  1.4 \\
234-096  & NOT   & H51  & -19.8 & E/S0                    & 07 11 38.2  & +46 56 09 & 1.0 & \ldots & 18 & 13.96 &   7(1) &  1.2(0.3) &  0.8 \\
234-102  & NOT   & H51  & -21.0 & Sb:                     & 07 12 51.0  & +49 00 21 & 4.0 & 153 & 58 & 13.48 &  10(2) &  1.1(0.3) &  1.2 \\
234-103  & NOT   & H51  & -19.9 & Sa:                     & 07 12 56.0  & +49 45 53 & 1.2 & 152 & 26 & 13.69 &   6(2) &  0.9(0.2) &  0.8 \\
234-107  & JKT94 & H668 & -19.8 & Sc                      & 07 13 47.6  & +50 15 26 & 2.0 &  56 & 21 & 14.10 &  14(2) &  2.7(0.3) &  1.7 \\
234-114  & NOT   & H51  & -19.5 & SAa:                    & 07 16 13.0  & +48 16 19 & 1.0 & \ldots & 18 & 14.28 &   1(1) &  0.2(0.2) &  0.1 \\
234-115  & NOT   & H51  & -19.3 & \ldots                  & 07 16 52.3  & +49 52 29 & 2.0 & 148 & 24 & 14.57 &   3(1) &  0.9(0.2) &  0.3 \\
\multicolumn{2}{@{}l}{{\it Abell 779}} &&&&&&&&&&&& \\
180-060  & JKT98 & H671 & -19.7 & Sa:                     & 09 13 45.4  & +34 50 14 & 1.3 &  58 & 17 & 14.36 &   7(1) &  1.6(0.3) &  1.1 \\
\end{tabular}
\end{table*}

\addtocounter{table}{-1}
\begin{table*}
\caption{-- continued}


\begin{tabular}{@{}l@{\hspace{0.9em}}l@{\hspace{0.9em}}l@{\hspace{0.9em}}r@{\hspace{0.9em}}
l@{\hspace{0.9em}}c@{\hspace{0.9em}}c@{}c@{}c@{}c@{}c@{}
c@{}c@{}c@{}}

\hline
 CGCG & Run & H$\alpha$ & $M_{\rm B}$ & Type & \multicolumn{2}{c}{R.A. (2000) Dec.} & 
\multicolumn{1}{c}{a/b} & \multicolumn{1}{c}{PA} & 
\multicolumn{1}{c}{$r_{24}$} & \multicolumn{1}{c}{$m_{\rm R}$} & 
\multicolumn{1}{c}{H$\alpha$ flux} & \multicolumn{1}{c}{H$\alpha$ EW} & 
\multicolumn{1}{c}{SFR} \\
 && filter &&& (h \hspace{0.5em} m \hspace{0.5em} s) & 
(\degr \hspace{0.5em} \arcmin \hspace{0.5em} \arcsec) && \multicolumn{1}{c}{($^{\circ}$)} & \multicolumn{1}{c}{(\arcsec)} &&
\multicolumn{1}{@{}c@{}}{($10^{-17} {\rm W} {\rm m}^{-2}$)} & \multicolumn{1}{c}{(nm)} & 
\multicolumn{1}{@{}c}{(M$_{\odot}$yr$^{-1}$)} \\ \hline
181-007  & NOT   & H51  & -19.5 & SA:a:                   & 09 17 56.2  & +34 30 34 & 1.2 & 139 & 19 & 14.65 &   4(1) &  1.3(0.3) &  0.7 \\
181-012  & NOT   & H51  & -20.1 & Sa:                     & 09 18 34.0  & +34 17 38 & 3.0 &  21 & 34 & 14.46 &   0(1) &  0.1(0.2) &  0.0 \\
181-013  & JKT94 & H671 & -20.6 & Sb:                     & 09 18 35.7  & +34 33 11 & 4.7 & 153 & 43 & 14.13 &   3(1) &  0.6(0.2) &  0.5 \\
181-016  & NOT   & H51  & -19.8 & SBa                     & 09 19 17.4  & +34 00 29 & 1.8 &  35 & 30 & 13.92 & \ldots & \ldots & \ldots \\
181-017  & NOT   & H53  & -20.5 & S:a:                    & 09 19 22.4  & +33 44 34 & 3.3 &  85 & 49 & 13.69 & \ldots & \ldots & \ldots \\
181-023  & JKT94 & H671 & -20.5 & S                       & 09 19 41.4  & +33 44 17 & 4.3 & 111 & 41 & 14.20 &   4(1) &  0.9(0.2) &  0.7 \\
181-030  & JKT94 & H671 & -20.1 & SB:b                    & 09 20 36.9  & +33 04 29 & 2.7 &   3 & 28 & 14.24 &   6(1) &  1.2(0.3) &  0.9 \\
181-032  & JKT98 & H674 & -20.4 & SBb                     & 09 20 52.7  & +35 22 06 & 1.1 &  98 & 30 & 13.31 &  20(3) &  1.8(0.3) &  3.4 \\
181-045  & NOT   & H51  & -20.0 & S:b:                    & 09 27 16.0  & +34 25 38 & 2.5 & 174 & 27 & 14.49 &   2(1) &  0.4(0.2) &  0.3 \\
\multicolumn{2}{@{}l}{{\it Abell 1367}} &&&&&&&&&&&& \\
126-074  & JKT94 & H671 & -20.2 & SBb pec                 & 11 31 01.9  & +20 28 21 & 1.7 &  88 & 24 & 13.54 &  35(3) &  3.9(0.4) &  5.0 \\
126-075  & JKT94 & H671 & -19.8 & SB                      & 11 31 03.7  & +20 14 08 & 2.0 &  57 & 23 & 13.59 & \ldots & \ldots & \ldots \\
126-100  & NOT   & H51  & -20.3 & SB:ab                   & 11 34 53.5  & +20 29 17 & 3.2 &  67 & 41 & 14.39 &  11(1) &  3.0(0.3) &  1.6 \\
126-104  & JKT98 & H671 & -19.8 & SAB...pec               & 11 35 35.0  & +20 30 19 & 2.3 & 116 & 25 & 14.67 &  16(1) &  5.1(0.4) &  2.3 \\
127-002  & NOT   & H53  & -19.6 & Sa pec                  & 11 36 53.2  & +21 00 15 & 1.4 &  47 & 21 & 14.40 &   8(1) &  2.1(0.3) &  1.2 \\
 97-026  & JKT94 & H671 & -20.7 & SBa pec                 & 11 36 54.4  & +19 58 15 & 1.8 &   8 & 27 & 13.43 &  70(5) &  7.1(0.5) & 10.1 \\
 97-027  & JKT94 & H671 & -20.4 & SB:a pec                & 11 36 54.2  & +19 59 50 & 1.8 &  33 & 28 & 13.75 &  15(2) &  2.0(0.3) &  2.1 \\
 97-033  & JKT94 & H671 & -19.9 & S(B)a                   & 11 37 36.0  & +20 09 49 & 1.8 &  88 & 27 & 14.10 &   5(1) &  1.0(0.2) &  0.8 \\
127-012A & NOT   & H53  & -21.2 & Sc:                     & 11 37 53.9  & +21 58 53 & 3.4 & 106 & 72 & 13.19 &   6(3) &  0.5(0.2) &  0.9 \\
127-012B & NOT   & H53  & -19.6 & SAb                     & 11 37 55.0  & +21 59 08 & 1.3 &   1 & 19 & 14.30 &  18(2) &  4.3(0.4) &  2.6 \\
127-016  & NOT   & H53  & -20.1 & Sa:                     & 11 39 16.2  & +20 25 38 & 2.8 &  71 & 36 & 14.02 &   2(1) &  0.3(0.2) &  0.2 \\
 97-041  & NOT   & H51  & -19.7 & SAa:                    & 11 39 24.5  & +19 32 04 & 1.3 &  78 & 22 & 14.25 &   6(1) &  1.3(0.3) &  0.8 \\
127-025A & JKT98 & H671 & -20.5 & Sab: pec                & 11 40 44.2  & +22 25 46 & 1.0 & 113 & 27 & 13.24 &  23(3) &  1.9(0.3) &  3.3 \\
127-025B & JKT98 & H671 & -19.8 & S:                      & 11 40 44.5  & +22 26 48 & 2.5 &  56 & 29 & 13.97 &  21(2) &  3.4(0.3) &  2.9 \\
 97-062  & JKT94 & H674 & -19.8 & Sa: pec                 & 11 42 14.8  & +19 58 35 & 2.7 &  55 & 26 & 14.67 &   7(1) &  2.3(0.3) &  1.0 \\
 97-064  & JKT94 & H668 & -19.5 & SA0                     & 11 42 14.6  & +20 05 52 & 2.5 & 140 & 28 & 14.52 & \ldots & \ldots & \ldots \\
 97-068  & JKT94 & H668 & -20.6 & SBc pec                 & 11 42 24.5  & +20 07 10 & 1.6 & 104 & 33 & 13.53 &  26(3) &  2.9(0.3) &  3.7 \\
 97-072  & JKT98 & H671 & -20.4 & SBab                    & 11 42 45.2  & +20 01 57 & 1.8 & 114 & 33 & 13.74 &   9(2) &  1.2(0.3) &  1.3 \\
 97-073  & JKT94 & H674 & -19.4 & SAcd: pec               & 11 42 56.4  & +19 57 58 & 1.0 & \ldots & 18 & 14.93 &  18(1) &  7.4(0.5) &  2.6 \\
 97-079  & JKT98 & H671 & -19.3 & S: pec                  & 11 43 13.4  & +20 00 17 & 1.6 & 108 & 17 & 15.38 &  24(1) & 14.8(0.9) &  3.5 \\
 97-087  & JKT98 & H671 & -21.6 & Sd pec                  & 11 43 49.1  & +19 58 06 & 6.3 & 134 & 67 & 13.28 &  80(5) &  7.0(0.5) & 11.4 \\
 97-091  & JKT98 & H674 & -20.5 & SABb                    & 11 43 59.0  & +20 04 37 & 1.4 &  54 & 30 & 13.58 &  27(3) &  3.1(0.3) &  3.9 \\
 97-092A & JKT98 & H671 & -19.2 & S0 pec                  & 11 43 58.2  & +20 11 06 & 1.6 &  14 & 15 & 15.19 &  10(1) &  5.3(0.4) &  1.5 \\
 97-093  & JKT94 & H668 & -19.9 & Sa:                     & 11 44 01.9  & +19 47 04 & 3.0 & 139 & 30 & 14.85 &  10(1) &  3.6(0.3) &  1.4 \\
 97-102A & NOT   & H51  & -19.9 & Sa                      & 11 44 17.2  & +20 13 24 & 1.5 & 135 & 26 & 13.76 &   6(2) &  0.9(0.2) &  0.9 \\
 97-114  & JKT94 & H674 & -19.2 & S0/a: pec               & 11 44 47.8  & +19 46 24 & 1.0 & \ldots & 13 & 15.03 &  14(1) &  6.1(0.4) &  2.0 \\
 97-120A & JKT94 & H668 & -20.9 & SAab                    & 11 44 49.2  & +19 47 42 & 2.0 &  41 & 44 & 12.94 &   9(4) &  0.6(0.2) &  1.3 \\
 97-121  & NOT   & H51  & -20.8 & SBb pec                 & 11 44 47.1  & +20 07 30 & 1.5 &  30 & 38 & 13.11 &  11(3) &  0.8(0.2) &  1.5 \\
 97-122  & JKT94 & H668 & -20.7 & Sb pec                  & 11 44 52.2  & +19 27 15 & 4.0 &  56 & 39 & 13.80 &  23(2) &  3.2(0.3) &  3.3 \\
 97-125  & JKT94 & H674 & -19.8 & Sa: pec                 & 11 44 54.8  & +19 46 35 & 1.2 &  69 & 25 & 14.14 &   8(1) &  1.5(0.3) &  1.1 \\
 97-129A & NOT   & H50  & -21.6 & SABb                    & 11 45 03.9  & +19 58 25 & 1.8 &  73 & 65 & 12.43 &  27(6) &  1.2(0.3) &  3.9 \\
127-045  & JKT98 & H671 & -20.6 & SAa                     & 11 45 05.9  & +20 26 18 & 1.2 &  40 & 30 & 13.31 &   6(3) &  0.6(0.2) &  0.9 \\
127-046  & JKT94 & H674 & -20.0 & SB:bc pec               & 11 45 05.7  & +21 24 42 & 2.2 & 128 & 28 & 14.28 &   9(4) &  2.0(0.8) &  1.3 \\
 97-129B & NOT   & H51  & -20.1 & S:...                   & 11 45 07.0  & +19 58 01 & 2.3 & 113 & 31 & 14.59 &   5(1) &  1.5(0.3) &  0.7 \\
 97-138  & NOT   & H50  & -19.6 & SAc                     & 11 45 44.7  & +20 01 52 & 1.0 & \ldots & 19 & 15.00 &  15(1) &  6.6(0.5) &  2.1 \\
127-049  & JKT98 & H671 & -20.1 & SBab                    & 11 45 48.8  & +20 37 43 & 3.3 &  65 & 33 & 14.50 &  16(1) &  4.4(0.4) &  2.3 \\
127-050  & NOT   & H51  & -20.7 & SBc                     & 11 45 55.6  & +21 01 32 & 1.1 & 179 & 39 & 13.24 &  15(3) &  1.4(0.3) &  2.2 \\
127-051A & NOT   & H53  & -19.6 & SB0/a                   & 11 45 59.9  & +20 26 20 & 1.8 & 124 & 25 & 14.11 &  14(2) &  2.9(0.3) &  2.1 \\
127-051B & NOT   & H53  & -19.3 & Sa pec                  & 11 45 59.5  & +20 26 50 & 2.0 & 150 & 19 & 14.74 &   6(1) &  2.2(0.3) &  0.9 \\
127-052  & JKT98 & H671 & -21.1 & SA0                     & 11 46 12.2  & +20 23 30 & 1.6 &   7 & 48 & 12.61 &   9(5) &  0.4(0.2) &  1.3 \\
127-055  & JKT98 & H671 & -19.5 & SAa                     & 11 46 46.7  & +21 16 17 & 1.0 & \ldots & 14 & 14.19 &  18(2) &  3.6(0.3) &  2.5 \\
 97-149  & NOT   & H51  & -19.7 & SAa:                    & 11 47 15.0  & +19 10 33 & 1.2 &  84 & 22 & 14.31 &   2(1) &  0.6(0.2) &  0.3 \\
 97-151  & NOT   & H51  & -19.8 & Sa:                     & 11 47 28.2  & +18 03 13 & 3.5 &  25 & 26 & 15.06 &   2(1) &  0.8(0.2) &  0.2 \\
 97-152  & NOT   & H51  & -20.1 & SBa:                    & 11 47 39.3  & +19 56 22 & 2.5 & 128 & 31 & 13.97 &   4(1) &  0.7(0.2) &  0.6 \\
127-056  & JKT94 & H671 & -20.1 & Sb:                     & 11 48 27.5  & +21 09 23 & 3.3 &  79 & 31 & 14.66 &   5(1) &  1.5(0.3) &  0.7 \\
127-062  & NOT   & H51  & -20.0 & SB:a                    & 11 49 30.3  & +21 02 37 & 2.0 & 175 & 31 & 14.07 &   3(1) &  0.5(0.2) &  0.4 \\
 97-160  & JKT94 & H671 & -19.4 & S0/a                    & 11 50 33.4  & +17 51 29 & 1.2 &  81 & 17 & 13.85 &   2(2) &  0.3(0.2) &  0.3 \\
127-067  & JKT94 & H671 & -19.5 & S? pec                  & 11 50 39.5  & +20 54 26 & 1.3 & 173 & 17 & 14.35 &  11(1) &  2.5(0.3) &  1.5 \\
127-068  & JKT94 & H671 & -19.7 & S-Irr                   & 11 50 52.7  & +21 10 08 & 1.5 & 153 & 18 & 13.94 &  19(2) &  3.1(0.3) &  2.8 \\
127-071  & JKT94 & H671 & -19.4 & S pec                   & 11 50 55.4  & +21 08 43 & 1.7 &  75 & 14 & 14.45 &  24(2) &  6.2(0.5) &  3.5 \\
 97-168  & JKT98 & H674 & -20.1 & Sbc:                    & 11 51 48.4  & +19 21 29 & 4.5 & 106 & 28 & 15.63 &   5(0) &  3.7(0.3) &  0.7 \\
 97-169  & NOT   & H51  & -20.7 & Sc                      & 11 51 52.5  & +18 32 46 & 9.0 & 151 & 46 & 15.73 &   4(0) &  3.7(0.3) &  0.6 \\
 97-172  & NOT   & H53  & -19.7 & SABb:                   & 11 52 14.5  & +18 39 03 & 2.0 & 165 & 22 & 15.25 &   4(1) &  2.3(0.3) &  0.6 \\
\end{tabular}
\end{table*}

\addtocounter{table}{-1}
\begin{table*}
\caption{-- continued}


\begin{tabular}{@{}l@{\hspace{0.9em}}l@{\hspace{0.9em}}l@{\hspace{0.9em}}r@{\hspace{0.9em}}
l@{\hspace{0.9em}}c@{\hspace{0.9em}}c@{}c@{}c@{}c@{}c@{}
c@{}c@{}c@{}}

\hline
 CGCG & Run & H$\alpha$ & $M_{\rm B}$ & Type & \multicolumn{2}{c}{R.A. (2000) Dec.} & 
\multicolumn{1}{c}{a/b} & \multicolumn{1}{c}{PA} & 
\multicolumn{1}{c}{$r_{24}$} & \multicolumn{1}{c}{$m_{\rm R}$} & 
\multicolumn{1}{c}{H$\alpha$ flux} & \multicolumn{1}{c}{H$\alpha$ EW} & 
\multicolumn{1}{c}{SFR} \\
 && filter &&& (h \hspace{0.5em} m \hspace{0.5em} s) & 
(\degr \hspace{0.5em} \arcmin \hspace{0.5em} \arcsec) && \multicolumn{1}{c}{($^{\circ}$)} & \multicolumn{1}{c}{(\arcsec)} &&
\multicolumn{1}{@{}c@{}}{($10^{-17} {\rm W} {\rm m}^{-2}$)} & \multicolumn{1}{c}{(nm)} & 
\multicolumn{1}{@{}c}{(M$_{\odot}$yr$^{-1}$)} \\ \hline
 97-174  & NOT   & H53  & -19.8 & Sc:                     & 11 52 43.9  & +18 36 49 & 2.0 & 132 & 24 & 15.15 &   4(1) &  2.1(0.3) &  0.6 \\
 97-180  & JKT94 & H671 & -19.4 & Sab                     & 11 54 13.9  & +20 01 39 & 2.3 & 123 & 17 & 14.82 &   9(1) &  3.2(0.3) &  1.3 \\
127-072  & NOT   & H51  & -20.8 & SABbc                   & 11 51 01.1  & +20 23 57 & 1.0 & \ldots & 35 & 13.30 &  17(3) &  1.6(0.3) &  2.5 \\
127-073  & NOT   & H51  & -20.5 & SBab                    & 11 51 02.2  & +20 47 59 & 1.0 & \ldots & 34 & 13.35 &   8(2) &  0.8(0.2) &  1.1 \\
127-082  & NOT   & H51  & -20.5 & SAc                     & 11 51 59.8  & +21 06 30 & 1.3 & 121 & 28 & 13.61 &  21(2) &  2.6(0.3) &  3.0 \\
127-083  & NOT   & H51  & -20.0 & Sa                      & 11 52 19.9  & +21 06 08 & 1.3 & 153 & 22 & 13.83 &   6(2) &  1.0(0.2) &  0.9 \\
127-085  & JKT94 & H671 & -20.0 & Sa                      & 11 52 30.5  & +20 37 32 & 3.0 &  73 & 28 & 14.35 &   4(1) &  1.0(0.2) &  0.6 \\
127-090  & NOT   & H51  & -20.6 & SBa                     & 11 52 56.6  & +20 28 45 & 1.0 & \ldots & 31 & 12.98 &   4(3) &  0.2(0.2) &  0.5 \\
127-095  & JKT98 & H668 & -21.2 & SBb                     & 11 53 20.3  & +20 45 06 & 1.2 &  97 & 44 & 12.60 &  41(6) &  1.9(0.3) &  5.8 \\
127-096  & NOT   & H51  & -19.9 & SAa                     & 11 53 20.6  & +21 01 18 & 1.2 &  85 & 23 & 13.84 &   5(2) &  0.7(0.2) &  0.7 \\
127-100  & NOT   & H51  & -20.5 & SBab                    & 11 53 59.7  & +20 34 21 & 1.4 &  29 & 32 & 13.49 &   7(2) &  0.8(0.2) &  1.0 \\
\multicolumn{2}{@{}l}{{\it Abell 1656}} &&&&&&&&&&&& \\
159-101  & JKT98 & H674 & -19.4 & Irr$^{\dagger}$         & 12 52 48.9  & +27 24 06 & 1.3 &  89 & 14 & 14.95 &  14(1) &  5.7(0.4) &  2.9 \\
159-102  & JKT98 & H671 & -20.9 & Sab$^{\dagger}$         & 12 52 53.6  & +28 22 16 & 2.5 &  32 & 36 & 13.30 &  35(4) &  3.1(0.3) &  5.9 \\
160-020  & JKT98 & H668 & -19.2 & Sa$^{\dagger}$          & 12 56 06.1  & +27 40 40 & 1.4 & 171 & 13 & 14.92 &  26(2) & 10.3(0.6) &  4.4 \\
160-025  & NOT   & H51  & -21.1 & SBa                     & 12 56 27.8  & +26 59 15 & 1.3 & 150 & 36 & 12.81 &   3(4) &  0.2(0.2) &  0.6 \\
160-026  & JKT98 & H674 & -19.5 & S0/a:                   & 12 56 28.5  & +27 17 28 & 1.4 &  66 & 17 & 15.31 &  11(1) &  6.2(0.5) &  1.8 \\
160-031  & NOT   & H51  & -20.1 & Sa$^{\dagger}$                      & 12 56 49.7  & +27 05 38 & 4.0 & 147 & 36 & 14.48 &   1(1) &  0.3(0.2) &  0.2 \\
160-033  & JKT98 & H668 & -19.6 & E$^{\dagger}$           & 12 56 51.2  & +26 53 56 & 1.0 & \ldots & 14 & 14.42 &   8(1) &  2.0(0.3) &  1.3 \\
160-055  & JKT98 & H671 & -21.2 & SB:ab                   & 12 58 05.6  & +28 14 33 & 3.2 & 151 & 41 & 13.53 &  39(3) &  4.3(0.4) &  6.6 \\
160-058  & JKT98 & H674 & -20.2 & S                       & 12 58 09.3  & +28 42 31 & 3.3 &  84 & 29 & 15.14 &   6(1) &  2.9(0.3) &  1.0 \\
160-064  & JKT98 & H674 & -19.5 & pec                     & 12 58 35.3  & +27 15 53 & 1.2 &  69 & 15 & 15.18 &  10(1) &  5.0(0.4) &  1.7 \\
160-067  & JKT98 & H674 & -19.3 & pec                     & 12 58 37.3  & +27 10 36 & 1.3 &  12 & 13 & 14.97 &  18(1) &  7.4(0.5) &  3.0 \\
160-068  & NOT   & H53  & -20.5 & (R')SA0-?$^{\dagger}$   & 12 58 35.2  & +27 35 47 & 1.2 &  83 & 24 & 13.24 &   5(3) &  0.4(0.2) &  0.8 \\
160-071  & JKT   & H671 & -19.4 & S0/a$^{\dagger}$        & 12 58 52.1  & +27 47 06 & 1.8 & 133 & 20 & 14.81 &  21(1) &  7.4(0.5) &  3.5 \\
160-072  & JKT   & H668 & -19.9 & S0/a                    & 12 58 48.5  & +27 48 37 & 3.9 &  53 & 28 & 14.84 &   1(1) &  0.2(0.2) &  0.1 \\
160-075  & NOT   & H54  & -19.4 & pec                     & 12 59 02.1  & +28 06 56 & 1.2 &  31 & 14 & 14.93 &  13(1) &  5.4(0.4) &  2.1 \\
160-078  & NOT   & H51  & -20.0 & E/S0:                   & 12 59 05.3  & +27 38 40 & 1.1 &  79 & 20 & 14.34 &   6(1) &  1.4(0.3) &  1.0 \\
160-099  & JKT98 & H668 & -19.3 & Sa:                     & 12 59 40.1  & +28 37 51 & 1.0 & \ldots & 15 & 15.24 &   8(1) &  4.1(0.4) &  1.3 \\
160-127  & JKT98 & H674 & -19.3 & pec                     & 13 00 33.7  & +27 38 16 & 1.4 &  65 & 13 & 15.38 &   7(1) &  4.4(0.4) &  1.2 \\
160-130  & JKT98 & H674 & -20.4 & pec:                    & 13 00 38.0  & +28 03 27 & 3.3 & 151 & 30 & 14.62 &  13(3) &  3.9(0.9) &  2.2 \\
160-132  & JKT98 & H674 & -20.8 & S                       & 13 00 39.7  & +29 01 10 & 2.0 &  56 & 36 & 13.61 &  14(2) &  1.7(0.3) &  2.4 \\
160-139  & NOT   & H51  & -20.7 & SB:ab                   & 13 00 48.8  & +28 09 30 & 1.1 &  39 & 31 & 13.15 &   5(3) &  0.4(0.2) &  0.9 \\
160-147  & NOT   & H51  & -21.9 & SBa                     & 13 01 26.1  & +27 53 10 & 1.1 &   2 & 67 & 12.06 &  22(8) &  0.7(0.2) &  3.7 \\
160-148A & JKT98 & H671 & -20.3 & S pec                   & 13 01 25.3  & +29 18 50 & 1.7 & 125 & 27 & 13.85 &  19(2) &  2.8(0.3) &  3.1 \\
160-148B & JKT98 & H671 & -20.3 & S pec                   & 13 01 24.5  & +29 18 30 & 1.0 & \ldots & 27 & 12.91 & \ldots & \ldots & \ldots \\
160-150  & JKT98 & H674 & -19.8 & S pec                   & 13 01 25.1  & +28 40 37 & 1.5 & 110 & 19 & 14.70 &   2(1) &  0.5(0.2) &  0.3 \\
160-154  & JKT98 & H671 & -20.7 & Sab                     & 13 01 43.4  & +29 02 40 & 3.0 &  85 & 33 & 14.23 &   8(1) &  1.8(0.3) &  1.4 \\
160-156  & JKT98 & H671 & -19.1 & SA0$^{\dagger}$         & 13 02 00.2  & +27 46 58 & 2.0 &  52 & 14 & 14.72 &   2(1) &  0.5(0.2) &  0.3 \\
160-158  & JKT98 & H671 & -19.6 & S0 pec?$^{\dagger}$     & 13 02 07.9  & +27 38 54 & 1.4 &  76 & 16 & 14.67 &  12(1) &  3.7(0.3) &  2.0 \\
160-159  & NOT   & H53  & -20.9 & Sa:                     & 13 02 04.2  & +29 15 12 & 4.5 & 156 & 49 & 13.47 &   2(2) &  0.2(0.2) &  0.3 \\
160-160  & JKT98 & H674 & -19.6 & pec                     & 13 02 12.8  & +28 12 53 & 1.7 & 150 & 18 & 14.84 &  11(1) &  4.0(0.4) &  1.9 \\
160-169  & JKT98 & H668 & -19.4 & S$^{\dagger}$           & 13 03 05.9  & +26 31 52 & 1.5 & 176 & 16 & 14.79 &   5(1) &  1.7(0.3) &  0.8 \\
160-176A & NOT   & H51  & -22.1 & Sab                     & 13 03 49.9  & +28 11 09 & 3.0 &  88 & 71 & 12.69 & \ldots & \ldots & \ldots \\
160-179  & JKT98 & H668 & -19.8 & S: pec                  & 13 04 26.6  & +27 18 16 & 1.5 &  65 & 21 & 14.98 &  16(1) &  6.5(0.5) &  2.6 \\
160-180  & JKT98 & H674 & -19.3 & pec                     & 13 04 22.8  & +28 48 39 & 1.3 & 177 & 12 & 15.39 &  12(1) &  7.2(0.5) &  2.0 \\
160-191  & JKT98 & H668 & -20.5 & pec                     & 13 06 38.1  & +28 50 53 & 1.8 & 169 & 32 & 14.60 &  23(2) &  6.9(0.5) &  3.9 \\
160-193  & JKT98 & H674 & -19.3 & Sc+$^{\dagger}$         & 13 07 13.2  & +28 02 49 & 1.1 & 165 & 14 & 14.94 &   9(1) &  3.7(0.3) &  1.6 \\
\hline
\end{tabular}
\begin{minipage}{18.0cm}
\noindent Explanation of columns of Table \ref{gparams}.

\noindent Column 1. CGCG number (Zwicky et al. 1960--68).  The numbering of CGCG galaxies in field 160 
(Abell 1656) follows that of the SIMBAD data base, with an enumeration in order of increasing R.A. 

\noindent Column 2. Observing run: JKT92, JKT94, JKT97 -- Jacobus Kapteyn Telescope 
1992 (service data), December 1994 and November 1997 respectively.
NOT -- Nordic Optical Telescope (5--8 January 2005).

\noindent Column 3. H$\alpha$ filter used. (For transmission profiles, see Figures \ref{fig:trans}a and \ref{fig:trans}b.) 

\noindent Column 4. Absolute B magnitude (see section \ref{abmag})

\noindent Column 5. Galaxy type taken from Moss, Whittle \& Pesce (1998); Moss \& Whittle (2000, 2005). 
For galaxies for which no type is listed by these authors, the NED type is
given (marked by $^{\dagger}$ in the Table).

\noindent Columns 6 \& 7. R.A. and Declination (J2000) taken from NED.

\noindent Columns 8, 9 \& 10. The major-to-minor axis ratio, position angle, and 
24 mag arcsec$^{-2}$ isophotal radius respectively.

\noindent Column 11. R band apparent magnitude.  The typical error in this measurement is $\sim 0.02$
mag.

\noindent Columns 12 \& 13. H$\alpha$ flux and EW 
within the R band 24 mag arcsec$^{-2}$ isophote. (Error is given in parentheses.)

\noindent Column 14. Global star formation rate (see section \ref{starfr}).
\end{minipage}
\end{table*}

\subsection{Absolute B magnitudes}
\label{abmag}

Absolute B magnitudes for the sample galaxies were calculated from CGCG
photographic magnitudes.  These magnitudes were converted to the $B_{T}$
system following RC3 and corrected for Galactic absorption using extinction
maps of Schlegel et al. (1998) and the conversion of Cardelli et al.
(1998).  A correction was also applied for internal extinction, dependent
on the inclination of the galaxy, as detailed in RC3.  K-corrections were
taken from Poggianti (1987) for the mean sample cluster redshift $z \sim 0.02$.
Cluster distances were found from mean cluster recession velocities corrected
to the centre of the local group following de Vaucouleurs et al. (1976),
and assuming $H_{0} = 75$ km ${\rm s}^{-1}$ ${\rm Mpc}^{-1}$.  All sample
galaxies were assumed to be cluster members, and the appropriate cluster
distance was used to obtain the absolute magnitude.

\subsection{Isophotal radii and R magnitudes}
\label{Rmag}

Following the methods of Koopmann, Kenney \& Young (2001), an outer isophotal
radius, $r_{24}$, is defined at 24 mag ${\rm arcsec}^{-2}$ in the R
band. This provides a direct tracer of size or luminosity, and allows 
the normalisation of profiles of galaxies of different sizes or distances.

For the observed galaxies, the $r_{24}$ isophote was found by calculating
the mean surface brightness of the annulus formed by each successive pair 
of apertures in the R band growth curves; this is then set to be the local
surface brightness at the mean semi-major axis of each annulus.  Linear
interpolation between points gives the semi-major axis at which the local
surface brightness first drops below 24 mag ${\rm arcsec}^{-2}$.  

The apparent magnitude, $m_{R}$, corresponds to the integrated R band
light out to the $r_{24}$ isophote.  The uncertainty in this magnitude
arising from errors associated with sky and other background
subtraction and calibration is typically quite small ($\delta m \sim
0.02$).  This magnitude was taken as the global value for a galaxy in
the subsequent analysis.

It is to be noted that $m_{R}$, determined in the above manner,
underestimates the total luminosity from the galaxy, since light from
outer isophotes fainter than $\mu = 24$ mag ${\rm arcsec}^{-2}$ has
been omitted.  A study was made, by comparison with the H$\alpha$GS
field sample, to estimate how large is this effect.  H$\alpha$GS galaxy R
band magnitudes were measured in apertures set such that the enclosed
flux varied by less than 0.5\% over three consecutive points in the
growth curve. The total magnitude was compared with $m_{R}$, the
magnitude at $r_{24}$, calculated for the H$\alpha$GS sample in the
same way as for the cluster data.  Total R band magnitudes are
typically found to be $< 0.1$ mag brighter than those measured at
$r_{24}$.

\subsection{H$\alpha$ flux and EW}
\label{haflux}

The H$\alpha$ flux is the integrated H$\alpha$ flux out to the
$r_{24}$ isophote; the EW is this integrated flux divided by the
corresponding R band flux density to the same isophotal limit.  It is
to be noted that the EW normalises the star formation
rate to the underlying older stellar population, thus providing a
measure of the specific star formation rate (i.e. the star formation
rate per unit mass). Kennicutt (1998) found that only high mass ($> 10$ M$_{\sun}$) and therefore short-lived ($< 20$ Myr) stars made any significant contribution to integrated ionising flux, so EW is a measure of the
importance of current or very recent star formation relative to
the star formation history of the galaxy.

\begin{figure}
\includegraphics[width=87mm]{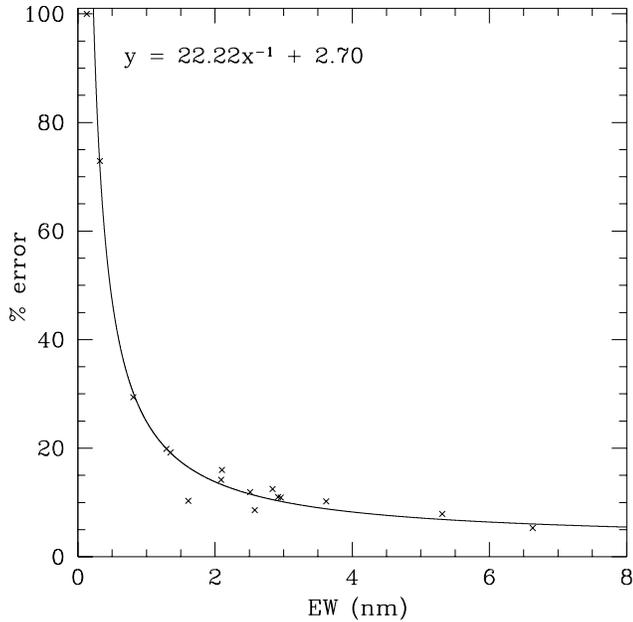}
\caption{Continuum scaling errors as a function of EW.}
\label{fig:cserrs}
\end{figure}

Errors in the H$\alpha$ flux and EW associated with sky and other
background subtraction and calibration are typically small ($\sim 3\%$
in flux; and $\sim 4\%$ in EW respectively). In contrast, the largest
single source of error for both flux and EW comes from the
determination of the scaling factor used to subtract the continuum
from the H$\alpha$ data.  The continuum scaling errors were found to
depend strongly on EW.  Figure \ref{fig:cserrs} shows the percentage error on
H$\alpha$ fluxes and EWs, due to a 1$\sigma$ change in continuum
scaling factor.  The points are well fit by a power law, $y = 22.22
x^{-1} + 2.70$ and this has been used in the calculation of individual
galaxy errors.  These errors range from 4\% for the highest EW objects
to $\sim 100\%$ for those with very low EWs, although the
uncertainties are less than 25\% for most galaxies within the sample.

A number of galaxies in the current sample are found in the literature
as part of other studies.  H$\alpha$ fluxes for 32 objects and EWs for
24 are given in Moss, Irwin \& Whittle (1988) (MWI88). A later paper
(Moss et al.  1998) (MWP98), based on similar observations in Abell
1367, includes further data for 18 galaxies.  Thirty objects in our
sample are included in an H$\alpha$ imaging survey by Gavazzi et
al. (1998) (GAV98), 29 in a large aperture photometry survey in the
Cancer, Abell 1367, and Coma clusters (Kennicutt, Bothun \& Schommer
1984) (KBS84), 15 in a deep H$\alpha$ survey of galaxies in Abell 1367
and Coma by Iglesias-P\'aramo et al. (2002) (IP02), a further 7 in a
study of spiral galaxies in the Coma and Hercules superclusters and
the Cancer cluster (Gavazzi, Boselli \& Kennicutt 1991) (GBK91), and
one object in a survey of star formation in spiral galaxies by
Romanishin (1990) (Rom90).

\begin{figure}
\includegraphics[width=87mm]{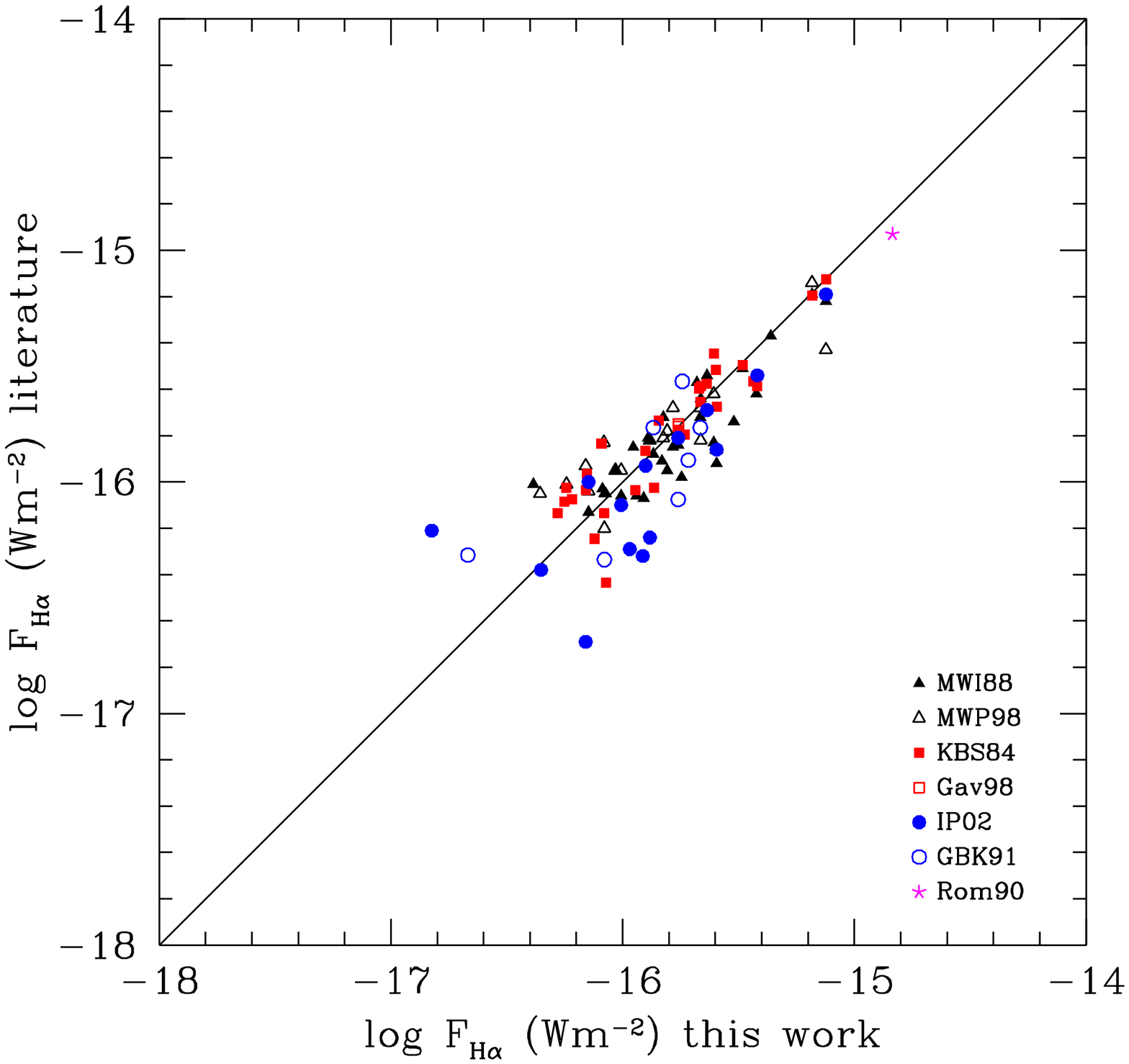}
\includegraphics[width=87mm]{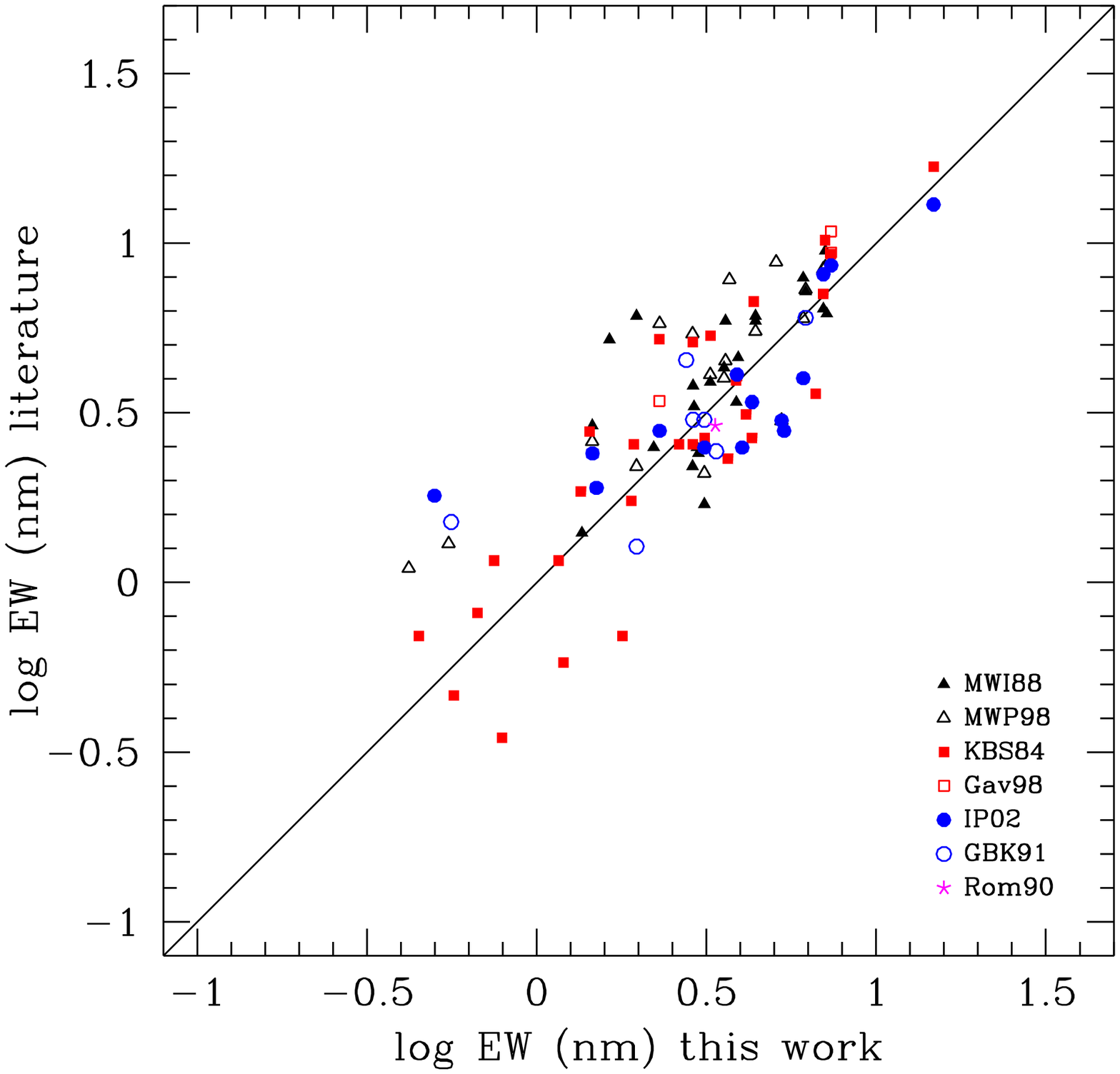}
\caption{Comparison of total H$\alpha$ fluxes (top) and EWs (bottom) obtained in this study with values in the literature.}
\label{fig:comp}
\end{figure}

Figure \ref{fig:comp} (top) shows a comparison between published
H$\alpha$ fluxes and those found in the present survey.  The flux
values taken from the literature are uncorrected for extinction
effects.  Therefore the Galactic extinction corrections applied to the
data in the present survey were removed for comparison purposes.  No
correction has been made for the satellite [NII] lines in any of the
comparison data; however the KBS84 and GBK91 fluxes and EWs have been
multiplied by 1.16 as suggested by Kennicutt et al. (1994) and Gavazzi
et al. (1998) to account for an overestimate of continuum flux in
these data, due to the telluric absorption feature near 6900 \AA\/ in
their red continuum side-band.  The solid line shows a one-to-one
correlation, which is a good fit to the data.  The rms scatter in
$\log F_{H\alpha}$ from 101 comparison measures is 0.18

A comparison of calculated EWs with 95 values from the literature is
shown in Figure \ref{fig:comp} (bottom). Again a one-to-one
correlation (solid line) shows good agreement with an rms scatter of
0.20 in $\log EW$.

No corrections have been made to account for the different aperture
sizes used in the comparison studies, or for the fact that the total
fluxes and EWs in this study are taken at $r_{24}$.  The true total
fluxes are therefore likely to be slightly larger than the literature
and current values shown here, as these apertures may not be detecting
all of the H$\alpha$ emission from the galaxies.  In a similar manner
as for the R magnitudes (see section \ref{Rmag} above), a study was made, by
comparison with the H$\alpha$GS field sample, to estimate how large is
this effect.  It was found that $r_{24}$ H$\alpha$ flux values are
typically $\sim 5\%$ lower than corresponding total values for
H$\alpha$GS galaxies.

\subsection{Star formation rates}
\label{starfr}

Cluster distances (see section \ref{abmag}) were used to convert H$\alpha$ +
[NII] fluxes into luminosities.  These latter were corrected for
internal extinction using a constant value of 1.1 mag (Kennicutt
1983). Following Kennicutt \& Kent (1983) a further correction was
made to account for the [NII] doublet that lies within the H$\alpha$
filter bandpasses. Corrected H$\alpha$ luminosities were then converted to star formation
rates (SFRs) using the relation,

$SFR(M_{\sun} yr^{-1}) = 7.94 \times 10^{-35} L_{H\alpha}(W)$

(Kennicutt, Tamblyn \& Congdon 1994), which assumes a Salpeter IMF
(Salpeter 1955) with masses ranging from 0.1 to 100 $M_{\sun}$.

Finally, it may be noted that the correction that was made to the
H$\alpha$ luminosity for contamination by the [NII] doublet depends
only on galaxy type, not on luminosity.  As is well known, lower
luminosity galaxies tend to be less metal-rich than their more
luminous counterparts (e.g. Zaritsky, Kennicutt \& Huchra 1994; Miller
\& Hodge 1996); this leads to the expectation of a lower
[NII]/H$\alpha$ ratio with decreasing galaxy luminosity, which is
indeed confirmed observationally (e.g. Jansen et al. 2000; Gavazzi et
al. 2004). However, the galaxies in the current dataset cover a
relatively narrow magnitude range, brighter than $M_{B} \sim -18.5$.
Furthermore, both Jansen et al. (2000) and Gavazzi et al. (2004) find
no significant variation in the [NII]/H$\alpha$ ratio within this
magnitude range. Thus neglect of any dependence of the [NII]/H$\alpha$
on galaxy luminosity for the correction to H$\alpha$ luminosities (and
hence SFRs) for the present galaxy sample appears to be justified.

\section{Completeness of the Objective Prism Survey}
\label{complet}

The current dataset is for a subsample of galaxies included in the
H$\alpha$ objective prism survey (OPS) undertaken by Moss and
colleagues (see Moss \& Whittle, 2000, 2005).  Moss et al. (1998)
estimated that ELG detection for the OPS was 90\% complete above 20
\AA  (2 nm) EW, and 17\% complete below this limit.  However these
estimates were based on a comparatively small sample of 35 galaxies,
for which photoelectric data were available.  By contrast, the current
CCD dataset includes a complete sample of Sa--Sc galaxies ($n=114$) in
six of the eight OPS clusters, of which $\sim$ 43\% ($n=46$) were detected
as emission-line galaxies (ELGs) by the OPS.  This sample may be used to 
provide a better determination of the completeness of the OPS, as follows.

\begin{figure}
\hspace{-5mm}
\vspace{-8mm}
\includegraphics[width=90mm]{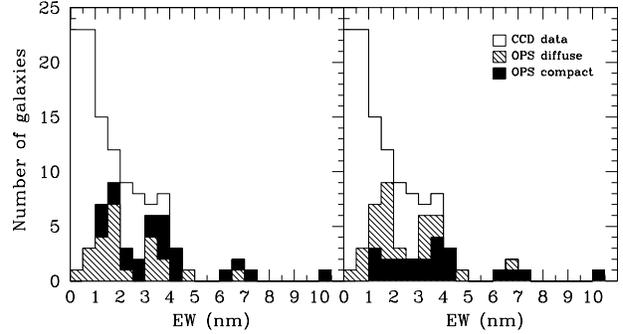}
\vspace{-35mm}
\caption{ Distribution of H$\alpha$ EWs for the complete Sa--Sc 
sample (open histogram) and those detected by the objective prism
survey as having diffuse (hashed histogram) or compact (solid
histogram) emission. The total shaded area shows the total number of
galaxies detected by the objective prism survey.}
\label{compew}
\end{figure}

\begin{figure}
\hspace{-5mm}
\vspace{-8mm}
\includegraphics[width=90mm]{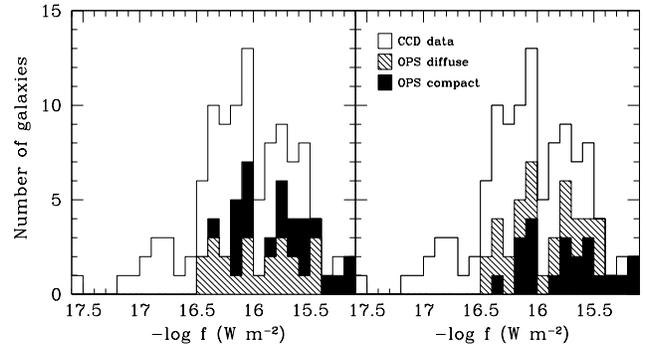}
\vspace{-35mm}
\caption{As Figure \ref{compew} for  H$\alpha$ flux values.
There are no objective prism survey (OPS) detections with $-log f >
16.5$.  Galaxies identified by the OPS as having diffuse emission
(hatched histogram) have a lower mean flux value than those classified
as compact (solid histogram).}
\label{fig:compf}
\end{figure}

The distribution of ELGs and non-ELGs for the Sa--Sc complete sample with
both H$\alpha$ EW and flux are shown in Figures \ref{compew} and 
\ref{fig:compf} respectively.  As is seen from Figure \ref{compew},
EW is a relatively poor discriminant between ELGs and non-ELGs: while 
63\% of galaxies are detected as ELGs above 2 nm, some 30\% of galaxies
are detected below this limit.

By contrast, H$\alpha$ flux appears as a rather better discriminant:
below a flux limit, $f = 3.2 \times 10^{-17} {\rm W} {\rm m}^{-2}$
($\log f = -16.5$), no galaxies were detected in emission by the OPS;
above this limit, the detection efficiency is 49\%, rising to 67\% for
galaxies with the strongest H$\alpha$ flux, $f \ge 1.6 \times 10^{-16}
{\rm W} {\rm m}^{-2}$ ($\log f \ge -15.8$).

\begin{figure}
\hspace{-5mm}
\vspace{-8mm}
\includegraphics[width=90mm]{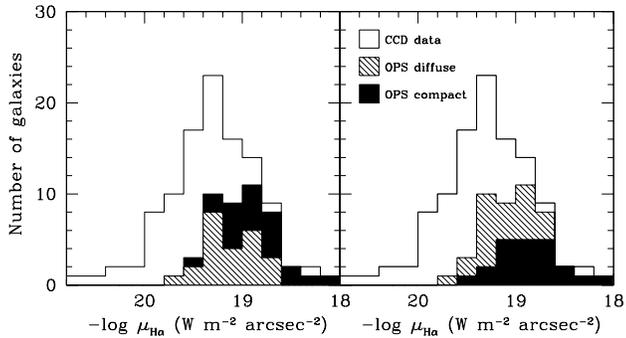}
\vspace{-35mm}
\caption{As Figure \ref{compew} for  H$\alpha$ surface brightness values.
Detections of galaxies by the objective prism survey (OPS) fall off
rapidly fainter than $-\log\mu_{H\alpha} = 19.4$. Galaxies identified
by the OPS as having compact emission (solid histogram) have a higher
mean surface brightness than those with diffuse emission (hatched
histogram).}
\label{fig:compsb}
\end{figure}

However the mean H$\alpha$ surface brightness appears to be a better
discriminant than either EW or flux.  In Figure \ref{fig:compsb} we
show distributions of ELGs and non-ELGs with mean H$\alpha$ surface
brightness, $\mu_{H\alpha} = f/A$ where $A$ is the area of the ellipse
centred on the galaxy bounded by the 24 mag ${\rm arcsec}^{-2}$
isophote. Below a limit, $\mu_{H\alpha} = 4 \times 10^{-20} {\rm W}
{\rm m}^{-2} {\rm arcsec}^{-2}$ ($\log \mu_{H\alpha} = -19.4$), only 4
out of 41 galaxies ($\sim$ 10\%) were detected as ELGs by the OPS;
above this limit, 63\% were detected, rising to 85\% above the limit
$1.25 \times 10^{-20} {\rm W} {\rm m}^{-2} {\rm arcsec}^{-2}$ ($\log
\mu_{H\alpha} = -19.9$).

In Figures \ref{compew}, \ref{fig:compf}, and \ref{fig:compsb} detected
ELGs are shown as {\it compact} or {\it diffuse} following their
classification by the OPS (for further details, see Moss \& Whittle
2000).  Compact ELGs have higher H$\alpha$ EW, flux, and surface
brightness than diffuse ELGs, but there is a broad overlap of all
these values between the two ELG types.  In particular, H$\alpha$ mean
surface brightness is not able to discriminate well between compact
and diffuse emission since this mean surface brightness is based on a
generally larger scale (viz. the $r_{24}$ isophote) than the smaller
scale features visually classified by the OPS. Further discussion of
the H$\alpha$ surface brightness distribution for individual galaxies,
and comparisons of these distributions between field and cluster
galaxies will be given in a subsequent paper (Bretherton et al.,
in preparation).

\section{Conclusions}
\label{concl}

H$\alpha$ and R band continuum CCD observations have been completed
for a sample of 227 CGCG galaxies associated with 8 low-redshift Abell
clusters, which were the subject of an objective prism survey (OPS) by
Moss and collaborators (Moss et al. 1998; Moss \& Whittle 2000,
2005). The sample galaxies were generally restricted to those with
velocities within 3$\sigma$ of the cluster mean, and known AGN have
been excluded. R band magnitudes, H$\alpha$ fluxes and EWs, and star
formation rates for the sample are listed in Table \ref{gparams}.

The dominant constraint on the detection efficiency of emission-line
galaxies (ELGs) by the OPS is shown to be H$\alpha$ surface
brightness. Detection of ELGs is 85\%, 70\%, and 50\% complete at the
mean surface brightness values of $1.25\times 10^{-19}$, $5.19\times
10^{-20}$, and $1.76\times 10^{-20}$ W m$^{-2}$ arcsec$^{-2}$ respectively,
where the mean H$\alpha$ surface brightness was measured within the R band
isophote of 24 mag ${\rm arcsec}^{-2}$ for the galaxy.

The present data, together with matched sets of data from a recent
H$\alpha$ galaxy survey of UGC galaxies within $v \le 3000$ km ${\rm
s}^{-1}$ (H$\alpha$GS, Shane 2002; James et al. 2004) will be used for
a forthcoming comparative study of R band and H$\alpha$ surface
photometry between cluster and field spirals (Bretherton et al., in
preparation).

\section*{Acknowledgments}

This research has made use of the NASA/IPAC Extragalactic Database
(NED) which is operated by the Jet Propulsion Laboratory, California
Institute of Technology, under contract with the National Aeronautics
and Space Administration. We thank Anna Hodgkinson for assistance with data 
preparation for the paper.

\end{document}